\documentclass[usenatbib]{mn2e}

\usepackage{epsfig}
\usepackage{amsmath}
\def\gsim{\mathrel{\raise0.35ex\hbox{$\scriptstyle >$}\kern-0.6em 
\lower0.40ex\hbox{{$\scriptstyle \sim$}}}}
\def\lsim{\mathrel{\raise0.35ex\hbox{$\scriptstyle <$}\kern-0.6em 
\lower0.40ex\hbox{{$\scriptstyle \sim$}}}}
\def\gs{\mathrel{\raise0.35ex\hbox{$\scriptstyle >$}\kern-0.6em 
\lower0.40ex\hbox{{$\scriptstyle \sim$}}}}
\def\ls{\mathrel{\raise0.35ex\hbox{$\scriptstyle <$}\kern-0.6em 
\lower0.40ex\hbox{{$\scriptstyle \sim$}}}}
\def\halpha{{\rm H$\alpha$}}
\def\ha {{\rm H$\alpha$}}
\def\oii{{\rm [O{\sc ii}]}}
\def\ewhanii{{\rm W$_\circ$(H$\alpha$+[N{\sc ii}])}}
\def\ewha{{\rm W$_\circ$(H$\alpha$)}}
\def\ewoii{{\rm W$_\circ$([O{\sc ii}])}}
\def\oiii{{\rm [O{\sc iii}]}}
\def\nii{{\rm [N{\sc ii}]}}
\def\znb{{\rm $(z^\prime-{\rm NB}_{912})$}}
\def\nb912{{\rm NB$_{912}$}}
\def\kmsmpc{{\,\rm km\,s^{-1}Mpc^{-1}}} 
\def\kms {{\,\rm km\,s^{-1}}} 
\def\zphot{{\,$z_{\rm phot}$}}
\def\lha{{\,L(\halpha+\nii)}}
\def\lshanii{{\,L$^\ast$(\halpha+\nii)}}
\def\lsha{{\,L$^\ast$(\halpha)}}
\def\lesssim{\mathrel{\hbox{\rlap{\hbox{\lower4pt\hbox{$\sim$}}}\hbox{$<$}}}}
\def\gtrsim{\mathrel{\hbox{\rlap{\hbox{\lower4pt\hbox{$\sim$}}}\hbox{$>$}}}}

\date{\today}
\title[\halpha\ imaging of Cl\,0024]{A Panoramic H$\alpha$ Imaging 
Survey of the $z$=0.4 cluster Cl\,0024.0+1652 with Subaru}

\author[Kodama et al.]
{Tadayuki Kodama$^1$, Michael L. Balogh$^{2,3}$,
Ian Smail$^2$, Richard G. Bower$^2$ \& \newauthor Fumiaki Nakata$^2$
\vspace*{6pt}\\
$^{1}$National Astronomical Observatory of Japan, Mitaka, Tokyo 181-8588,
Japan\\
$^{2}$Institute for Computational Cosmology, University of Durham, South Road,
Durham, DH1 3LE, UK\\
$^{3}$Present address: Department of Physics, University of Waterloo,
Waterloo, ON, Canada N2L 3G1
}

\begin{document}

\maketitle

\begin{abstract}
We employ panoramic, multicolour ($BRz^\prime$) and narrow-band 
H$\alpha$ imaging of the cluster Cl\,0024.0+1652 ($z=0.39$) from Subaru
covering a $\sim$ 30~arcmin field, to determine
cluster membership and star formation rates for a large
sample of galaxies across a wide field in the cluster, $\sim 10$\,Mpc.
We use photometric redshifts to identify cluster members, and
statistically correct for the residual field contamination using 
similar data from the Subaru Deep Field.  We detect over 500 
galaxies in narrow-band emission, with broad-band colours consistent with
them lying at $z\sim0.39$.
Using this sample we determine the H$\alpha$ luminosity
function within the cluster and find that its
form is approximately independent of local density, and 
is consistent with that seen in the intermediate redshift field population.  
This suggests that any density--dependent physical mechanisms which
alter the star formation rate must leave the H$\alpha$ luminosity
function unchanged; this is possible if the time-scale for star
formation to cease completely is short compared with a Hubble time.
Such short time-scale transformations are also supported by the presence
of a population with late-type morphologies but no detectable H$\alpha$
emission.
The fraction of blue galaxies, and the fraction of galaxies
detected in H$\alpha$, decreases strongly with increasing galaxy
density in a manner which is qualitatively similar to that seen at lower
redshifts.  This trend is significantly steeper than the trend with galaxy
morphology observed from a panoramic {\it Hubble Space Telescope} image
of this cluster; this suggests that the physical mechanisms responsible
for transformations in morphology and star formation rates may be
partially independent.  Finally, we compare our data with similar data
on clusters spanning a range of redshifts from
$0.2\lesssim z\lesssim0.8$ and find little evidence for a trend in the
total amount of star formation in clusters with redshift.  Although the
data can accommodate strong evolution, the scatter from cluster to
cluster at fixed redshift is of a comparable magnitude.
\end{abstract}


\section{Introduction}

The correlation between average stellar population age, star formation rate,
and galaxy environment at low redshift has recently been established
quantitatively, thanks to large redshift surveys
\citep{2dF_short,Sloan_sfr_short,Hogg03,Baldry03,2dfsdss}.  To first
order, the galaxy population is made up of two types: one
with blue stellar populations and active star formation, and a second,
quiescent one with redder colours.  Although the
average colour and star formation rate of the blue population is
approximately independent of environment, the relative fraction of
these galaxies decreases strongly as the local galaxy density increases
\citep{BB04,2dfsdss}.  This correlation exists on all density scales, even
in low--density regions well outside the cores of rich clusters.  

The next step is to measure how these relationships evolve with time.
\citet{BO78a} were the first to present convincing evidence that
galaxies in dense cluster cores were bluer in the recent past than they
are today.  However, the subsequent detection of a similar type of evolution
in the general field population \citep{L96} means that
we do not know how much of this evolution is due to external forces
(i.e.\ the environment) and how much is due to normal
galaxy evolution by, for example, the gradual consumption of available
cold gas.  To resolve this question requires measurements of the correlation
between galaxy stellar populations and environment at a series of
redshifts.  
Early attempts at this used sparsely sampled redshift surveys of
clusters to correlate galaxy star formation rates with environment
\citep[e.g.][]{B+97,F+98,P+99}.  These studies showed that,
even at $z\sim 0.5$ there are
few emission line galaxies within clusters, relative to the surrounding
field, even $\sim 1$ Mpc from the
cores.  In particular, they find relatively large populations of red
galaxies with mid/late-type disk morphologies but no signs of star formation;
such galaxies are rarely seen in the field \citep[][]{P+99}.  
Unfortunately, the limited fields of view of these studies meant it
was impossible to identify where the star formation in in-falling field
galaxies is terminated, or to relate this to any changes in their morphologies.
A key requirement for making progress is to trace the star-forming
population
within clusters across very large distances and hence a wide
range of environments, out to the turn-around
radius ($>5$\,Mpc).

Most observations of clusters at intermediate redshift have relied on
the \oii$\lambda3727$ emission line as a tracer of star formation.
As has frequently been pointed out, the
[O{\sc ii}] emission is severely affected by dust extinction and
metallicity; thus, another explanation of the 
apparently passive, red spiral galaxies in distant clusters could be
that they are more highly extincted or metal--poor than disk galaxies
in the field.  For this reason, 
H$\alpha$ has been more recently
used to study the distribution of star formation within
high-density environments, particularly at $z<0.5$ where the observed
wavelength lies in the visible spectrum.
One such programme which obtained a deeper, more complete measurement of the
star formation activity in clusters involved deep, ultraplex spectroscopic
surveys of narrow wavelength ranges, targetting H$\alpha$ emission in 
clusters at $z\sim 0.2$--0.3
\citep{C+01,A1689}.  This provides a very sensitive probe of the
H$\alpha$ luminosity function on $\sim 7$~arcmin (2--4\,Mpc) scales
within intermediate redshift clusters.  
These measurements have shown that, while the fraction of
galaxies detected in H$\alpha$ remains low even $\sim 1$ Mpc from the
cluster centre, at least the shape of the bright end of the H$\alpha$
luminosity function is approximately independent of environment.

One particularly powerful
approach to surveying H$\alpha$ within clusters
is to use narrow-band
imaging to trace H$\alpha$ emitting galaxies over very wide fields
in a narrow redshift slice \citep[e.g.][]
{Fujita_half,Umeda03,FZM}.  
Narrow--band imaging has the advantages that
it traces the two-dimensional distribution of instantaneous star
formation, yields the total star formation rate (SFR) independent of
aperture effects, and is complete (flux-limited) over very large fields
and hence a wide range of environments.     
However, these studies do not have the
high-resolution imaging over a comparable field, necessary to identify
the galaxies with late-type morphologies but low star formation rates
that might be indicative of a transformation in progress.

In this paper, we present multicolour (broad-band and narrow-band
H$\alpha$) Suprime-Cam observations over a
$\sim 9.6$ Mpc field around  the cluster Cl\,0024.0+1652
(Cl\,0024) at $z=0.395$.  We use the
broad-band colours to identify cluster members, and to trace the young
stellar population, while the narrow-band H$\alpha$ imaging follows the
instantaneous star formation rate.  
Importantly, a large {\it Hubble Space Telescope} ({\it HST}) mosaic,
consisting of 39 WFPC2 pointings, is available.  Analysis of these data
has shown  that the fraction
  of early type galaxies increases steeply from $\sim 1$ Mpc toward the
  cluster centre \citep{Treu03}.  The combination of these high
  resolution images with our new H$\alpha$ data provides a unique
  opportunity to study the effects of environment on morphology and
  star formation rate, separately.

Cl0024 is one of the original clusters studied by
\citet{BO78a,BO78b}, and later followed up spectroscopically
\citep{CN84,DG82,DG92,D+99,P+99,Czoske_cat}.  
From the lensing analysis of \citet{Kneib03}, the 
characteristic radius of the cluster is $R_{200}=1.7$ Mpc, and 
the total mass within this radius is $5.7\times10^{14}M_\odot$.  
The cluster core is dominated by red galaxies, with a secondary population of
spiral galaxies that are mostly unremarkable \citep{SDG}, though some show
evidence for recent starbursts \citep{K+97,P+99}, perhaps enshrouded in
dust \citep{Coia}.    The red galaxies follow
the fundamental plane relation \citep{vdF}, and are consistent with
passive evolution models \citep{Kodama+98,Barger98}.
Earlier {\it HST} imaging \citep{Smail} revealed a clear lack of S0 galaxies
  relative to nearby clusters  \citep{D+97}, which has been used as an
  argument that the S0 population is being formed within such
  clusters, between $z\sim 0.4$ and the present.  
Recent, wide-field spectroscopy of
a very large sample of cluster galaxies by \citet{Czoske_cat} has identified
substructures in the foreground and background of Cl\,0024. 
They find the velocity dispersion of the main
cluster component is only $\sim
  560$\,km\,s$^{-1}$, much less than earlier measurements of
  $\sim1200\kms$ \citep{DG92} which did not
  resolve the substructure.  
This spectroscopic sample of cluster galaxies is the largest available for
any cluster at $z\gs 0.2$ and enables us to perform a number of powerful
tests of the reliability of our narrow-band selected sample.

The present paper is organized as follows.
Our observations and data reduction are presented in \S~\ref{sec-obs},
and the measurements of H$\alpha$ emission and derived quantities are
described in \S~\ref{sec-halpha}.  The spatial distribution and
environmental variation in galaxy populations are shown in
\S~\ref{sec-results}.  We discuss the implications of these results for
galaxy evolution in general, in \S~\ref{sec-discuss}, and draw our
final conclusions in \S~\ref{sec-conc}.
Throughout this work we assume a cosmology with $\Omega_m=0.3$,
$\Omega_\Lambda=0.7$, $H_\circ=70\kmsmpc$.  At the redshift of Cl\,0024,
$z=0.395$, 1\,arcmin corresponds to 0.32\,Mpc, and the characteristic
luminosity $M^\ast$ is $z^\prime\sim19.2$.

\section{Observations and Data Reduction}\label{sec-obs}

Data were obtained with Suprime-Cam (0.202 arcsec per pixel, and
27~arcmin field of view) on the
Subaru telescope on
the night of September 6, 2002.
Images were obtained in three broad-band filters
($BRz^\prime$), and the narrow band filter
\nb912 ($\lambda_{\rm eff}$=9139~\AA, FWHM=134~\AA)
The net exposure times are 60~min in $B$, 88~min in $R$, 33~min in
$z^\prime$ and 180~min in \nb912.
We took two exposures in $z^\prime$, a deep one (4~min) and a shallow
one (1~min); the latter was required because the brightest
galaxies ($z^\prime<21$) are saturated on the deep image.
 The 5-$\sigma$ limiting magnitude of the $z^\prime$ images, on which our
selection is based, is 22.7 mag, corresponding to $M^\ast+3.5$ at the
cluster redshift.  The seeing was 0.7--1.0 arcsec in the $R$, $z'$ and
\nb912 bands, but worse ($\sim$1.0--1.3 arcsec) in the $B$-band.
The sky conditions were photometric, and the photometric zero-points were
calibrated based on the \citet{Landolt92} standard stars in $B$ and $R$;
the $z'$-band image was calibrated onto the SDSS system using the star
G24-9.
The data were reduced with the {\sc iraf} and {\sc nekosoft}
\citep{Y98} software packages, following standard
procedures of  bias subtraction and flat-fielding.  The latter is achieved
using supersky flats constructed  from
the median of  the dithered science frames.  We then mosaic the chips,
taking care to 
match the point-spread function (PSF) and
relative flux  calibration between them.

\begin{figure}
\leavevmode \epsfxsize=8.5cm \epsfbox{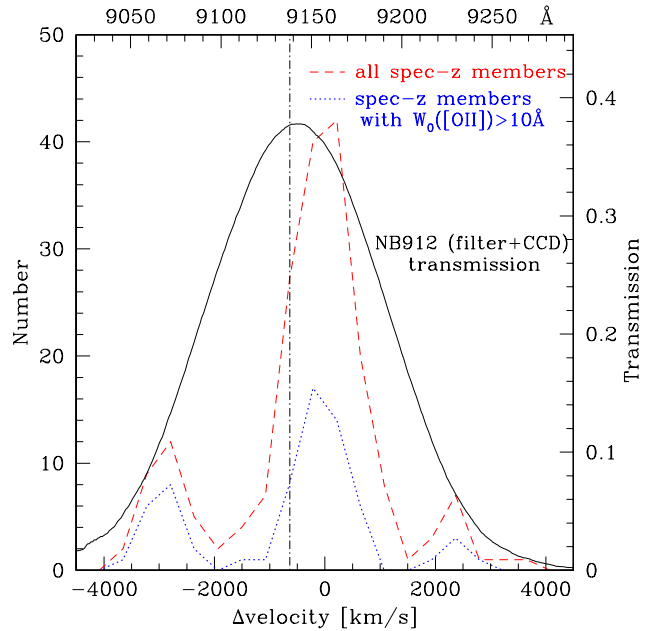}
\caption{The {\it
  dashed line} is the velocity distribution of spectroscopically
confirmed  cluster members in Cl\,0024, from \citet{Czoske_cat};
the {\it dotted curve} is the distribution
for those members with \oii$\lambda3727$ emission.
The top axis shows the corresponding wavelength of H$\alpha$.
The throughput of the  narrow-band filter \nb912, including the
response of the CCD, as a function of velocity is shown as the 
solid curve, according to the scale on the right axis.
\label{fig:velocity}}
\end{figure}

The relative transmission function of the narrow band filter \nb912
combined with the response of the CCD is shown
in Fig.~\ref{fig:velocity}. 
The effective wavelength and width are 9139~\AA\, and 134~\AA, respectively.
This response function is compared with the velocity
distribution of spectroscopically confirmed cluster members in our
target, Cl\,0024, from \citet{Czoske_cat}.
At the peak of the observed galaxy redshift
distribution, the H$\alpha$ emission line lies at $\lambda\sim9160$~\AA,
close to the maximum filter transmission.  The galaxy
redshift distribution is significantly narrower than \nb912, so we
have good sensitivity for all cluster members.  Note that the
foreground and background structures detected by \citet{Czoske_cat} lie
in regions of lower filter transmission, so our H$\alpha$ detection
limit will be brighter in these structures.

Galaxies are detected from the $z^\prime$ images, using the {\sc SExtractor}
software \citep{sextractor}.
We detect all objects with at least 9 connected pixels
(0.38 arcsec$^2$, equivalent to the area of the PSF) 
more than 2 $\sigma$ above the median sky.
The images in all filters are aligned by centroiding
stars throughout the field and fitting a stretch and shift.  The images
are each convolved with a Gaussian to mimic a seeing of 1.0 arcsec,
corresponding to that of our poorest quality data
in the $R$ and $z^\prime$-bands.
The seeing on the $B$-band images is slightly worse,
and these are convolved to 1.3 arcsec. 
As a result, the FWHM of the stars are not uniform
between the passbands; however, as we use a fixed 3~arcsec aperture
(corresponding to 16~kpc at the cluster redshift) when measuring galaxy
colours, this small mismatch in the seeing sizes produces a negligible
difference in the measured colours.
The magnitude {\sc mag\_best} is used as a measure of total magnitude.
All the magnitudes in this paper are given in the Vega-based system.

We will use the Subaru Deep Field \citep[SDF,][]{SDF1_short} as a control
field, to correct for residual background contamination after photometric
membership selection (e.g.\ Kodama et al.\ 2001).
This imaging was reduced using similar techniques,
and has an effective area of 667.94 sq. arcmin.  These observations
include the same broad band filters that we use, and also a similar
narrow-band filter, with a central wavelength of
9196~\AA\ and an effective FWHM of 132~\AA. 

\section{\halpha\ measurements and field subtraction}\label{sec-halpha}

\subsection{Continuum subtraction and detection criteria}\label{sec-cont}

\begin{figure}
\leavevmode \epsfysize=8.5cm \epsfbox{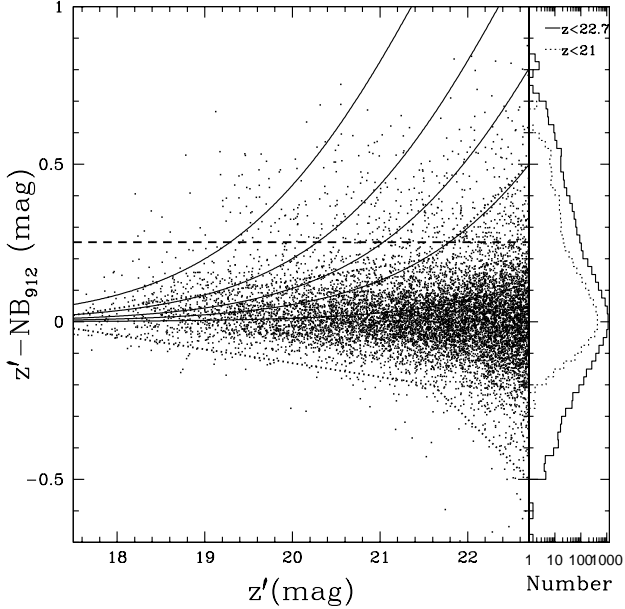}
\caption{The distribution of galaxies in \nb912--$z^\prime$ colour, as a
  function of $z^\prime$ magnitude for all galaxies in the photometric
  catalogue.  
The curved,  solid
    lines show lines of constant H$\alpha$ luminosity from 
\lha$=0.66, 3.3, 6.6, 13.2$ and $33 \times 10^{40}$ ergs~s$^{-1}$,
increasing towards higher $z^\prime$--\nb912.  These correspond to
SFR, in units of $M_\odot~$yr$^{-1}$,
of 0.1, 0.5, 1.0, 2.0 and 5.0, 
respectively.  The  {\it diagonal, dotted lines} show the envelope of
the observed scatter; the 753 galaxies that lie above the upper line are
secure emission line detections, although we estimate that $\sim 240$ are 
likely to be background galaxies on the basis of $BRz'$ colour-colour diagram
(see \S~3.2 and Fig.~7).
For galaxies with $z^\prime<21.8$, the sample of detections is
complete to  \ewhanii$=40$~\AA, shown as the {\it horizontal, dashed
line}.  The histograms on the right show the distribution of
\znb\ for all galaxies in the sample, brighter than $z^\prime=22.7$
(solid line) and $z^\prime=21$ (dotted line).  
\label{fig:znb}}
\end{figure}

Emission line fluxes are measured by comparing the aperture
magnitudes in the 
\nb912\ and $z^\prime$ bands, as shown  in Fig.~\ref{fig:znb}.  
We determine the zero-point of the \nb912\ filter by requiring the mean
\znb\ colour to be zero for stars and spectroscopic
non-members. 
Fig.~\ref{fig:znb}
shows that most detected galaxies have \znb$\sim 0$, with a scatter that
increases toward fainter magnitudes.  This is expected, because most
galaxies are in the foreground or background, and \halpha\ does not lie
within our filter bandpass; the scatter is primarily due to photometric
errors.  Since the equivalent widths of stellar H$\alpha$ absorption
lines in non-star forming galaxies
are typically only $\sim 2$~\AA\ \citep[e.g.][]{TragerVI},
we can assume the envelope of
galaxies at \znb$<0$ traces the intrinsic scatter due to
measurement uncertainty; this envelope is approximated by the
{\it dotted lines} in
Fig.~\ref{fig:znb}, given by $(z^\prime-{\rm NB}_{912}) >
0.044(z^\prime-21.6)+0.2$ for $z^\prime<21.6$,
and $(z^\prime-{\rm NB}_{912}) >
0.273(z^\prime-21.6)+0.2$ for $z^\prime\geq21.6$.  
There are 753 galaxies lying above this line that represent significant
detections; however, some of these will correspond to emission from
background galaxies, in lines other than H$\alpha$.  We make a
correction for these based on broad-band colours,
as described in \S~\ref{sec-photz}.

Since the \nb912\
filter has a Gaussian profile and is slightly offset from the 
wavelength of H$\alpha$ at the cluster redshift,
we must account for the lower transmission of the \nb912\ filter for
a typical cluster galaxy. 
We correct for this, in an average sense,
by multiplying our measured H$\alpha$ fluxes and equivalent widths (see
\S~\ref{sec-haflux}) by a factor $1.37$, compensating for the 
throughput averaged over the spectroscopic cluster members
in the Czoske sample, as shown in Fig.~\ref{fig:velocity}.  On an individual
galaxy basis, however, the H$\alpha$ flux for cluster members is
uncertain by $\sim 15$ per cent, due to the variable filter throughput
over $1000\kms$.  We have also checked our results for sensitivity to
colour terms; however we neglect this correction as it is small relative
to our photometric uncertainties.

\begin{figure}
\leavevmode \epsfysize=8.5cm \epsfbox{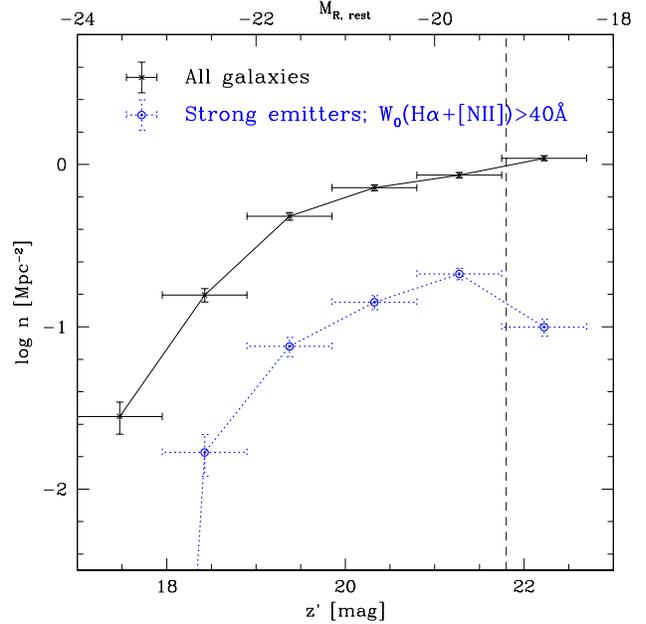}
\caption{The differential number count of galaxies in the $z^\prime$-band.
The solid line and the dotted line show all the galaxies and those with
strong emissions, respectively. The vertical dashed line represents
$z^\prime$=21.8 below which our detection of strong emitters become
incomplete due to large photometric errors.
\label{fig:ncount}}
\end{figure}

\subsection{Computation of \ewhanii\ and \halpha\ fluxes}\label{sec-haflux}

The emission lines \nii$\lambda\lambda6548,6583$ are adjacent to H$\alpha$,
and also contribute to
the emission line flux within \nb912.   Instead of making an uncertain
correction for this based on empirical calibrations
\citep[e.g.][]{T+99}, we choose to present most of our results for the
combined line flux, and only make a correction
when computing SFRs.  
To compute the rest frame equivalent width, \ewhanii, from \znb,
we need to correct for the effect of emission within the broad-band filter.
The conversion depends on the relative wavelength ranges of the $z^\prime$
filter
($\Delta \lambda_{BB} =1360$~\AA) and narrow-band filter 
($\Delta \lambda_{NB}=134$~\AA), and is given by:
\begin{equation}
{\rm W}_\circ({\rm H}\alpha+[{\rm NII}])=(1+z)^{-1}\times {\Delta \lambda_{\rm BB}
\Delta \lambda_{\rm NB} (r-1) \over \Delta\lambda_{\rm BB}- \Delta\lambda_{\rm NB}r},
\end{equation}
where $\log_{10}(r)=$\znb$/2.5$ and $z=0.395$ is the cluster redshift.
To calculate \halpha+\nii\ fluxes we
multiply \ewhanii\ by the continuum $z^\prime$ luminosity (on the Vega system,
$z^\prime=19.2$ corresponds to a flux of
$1.636\times10^{-17}$ ergs~s$^{-1}$cm$^{-3}$\AA$^{-1}$).  
In Fig.~\ref{fig:znb} we show the horizontal line
corresponding to \ewhanii=40~\AA; at all magnitudes brighter than
$z^\prime=21.8$, galaxies with
equivalent widths greater than this limit are significantly detected in
emission
(see also Fig.~\ref{fig:ncount}).
We will therefore use this magnitude limit when considering
the fraction of galaxies with \ewhanii$>40$~\AA, as in \S~\ref{sec-density}.

\begin{figure}
\leavevmode \epsfysize=8.5cm \epsfbox{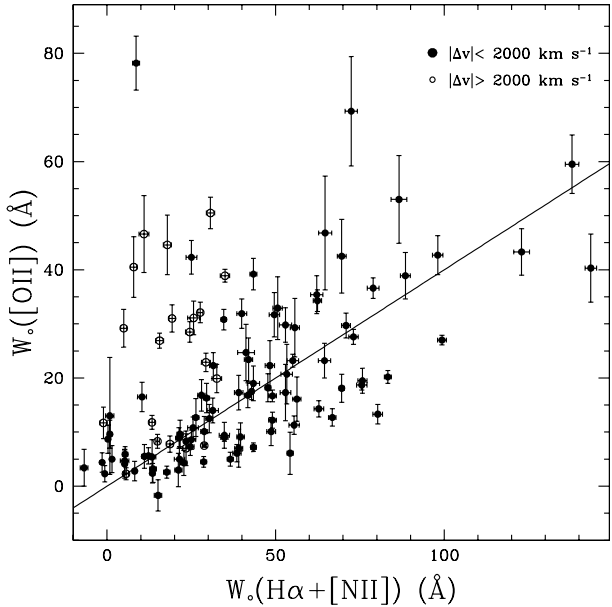}
\caption{The correlation between our measured
\ewhanii\ and the spectroscopically measured \ewoii\ from
\citet{Czoske_cat}.  {\it Solid circles} are galaxies with redshifts
within $2000\kms$ of the peak of filter transmission function, while
{\it open circles} are galaxies with velocity
differences greater than this.  The {\it solid line} is the local relation,
from \citet{K92}.  
\label{fig:haoii}}
\end{figure}

We then
compute total luminosities, \lha, using a luminosity distance of 2147 Mpc,
appropriate for our cosmological model.
Finally, we compute SFRs assuming the conversion of \citet{K+94},
correcting for 30\% \nii\ emission \citep{T+99,Jansen} and an 
extinction at \halpha\ of 1 magnitude \citep{K+94}.  Both of these
corrections are uncertain, and only accurate in an average sense,
for reasonably bright galaxy populations ($\gtrsim M^\ast+1$) with
metallicities above $\sim 0.5$ solar \citep{Brinchmann03,KGJ}.

In Fig.~\ref{fig:haoii} we show the correlation between our measured
equivalent widths, \ewhanii, and the \oii\ equivalent widths from
\citet{Czoske_cat}, for spectroscopically confirmed cluster members.
The correlation is in reasonable agreement with
the local relation from \citet{K92}, with considerable scatter.  Some
of this scatter is due to the $\sim$15 per cent uncertainty on the \nb912\
filter throughput.  The \ewha\ for galaxies with redshifts $>2000\kms$ from
the peak of the \nb912\ filter
($\lambda=9140$~\AA, corresponding to $z=0.393$ for H$\alpha$) are 
underestimated by typically a
factor of $2\times$, due to the lower filter transmission.  These
galaxies are shown as the open circles in Fig.~\ref{fig:haoii},
and all have low \ewha\ for their measured \ewoii.
From Fig.~\ref{fig:velocity} it is evident that these galaxies correspond to
the dynamically distinct foreground and background clumps, and not the
main body of the cluster.

\begin{figure}
\leavevmode \epsfysize=8.5cm \epsfbox{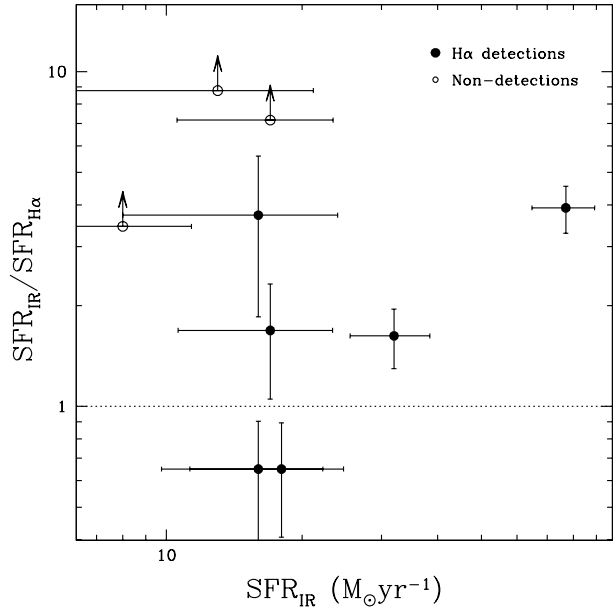}
\caption{The ratio of the SFR estimated from mid-infrared {\it
ISO}/CAM observations \citep{Coia} to that measured from our H$\alpha$
photometry.  The solid circles are H$\alpha$ detections, and the
open circles are non-detections.  Four ISO--detected galaxies are not
included in this comparison for reasons discussed in the
text, and thus do not appear on the plot.
\label{fig:Coia}
}
\end{figure}

Recently, \citet{Coia} presented {\it Infrared Space Observatory}
({\it ISO})/CAM mid-infrared observations of the
central $\sim$ 2~arcmin of Cl\,0024 at 15$\mu$m.  They
identify thirteen mid-infrared sources which 
are confirmed cluster members (within the main cluster, not the foreground
and background substructures), and they are able to derive
star formation rates for twelve of them. Two of these detections are not in
our catalogue because of corrupted photometry in our image, while a third
is excluded because $z_{\rm phot}$ is outside our membership limits,
and $z_{\rm spec}$ is flagged as uncertain (see \S~\ref{sec-photz}).
Of the remaining 9 galaxies, 6 are detected in H$\alpha$; we compare
the SFR estimated from H$\alpha$ (assuming 1 mag extinction) with
that estimated from the mid-infrared, in Fig.~\ref{fig:Coia}.
For all of the  {\it ISO} sources that we detect in emission, the
H$\alpha$--derived SFRs are within a factor of $\sim$2--3 of the
mid-infrared estimates.  For the three remaining galaxies,
the infrared SFRs are larger than those determined from H$\alpha$,
by at least a factor $\sim$3--10.  This discrepancy
is likely due to a wide range of extinctions among strongly star--forming
galaxies \citep{Hopkins01,Hopkins03,Afonso}, and implies that the
extinction for these galaxies is $\gtrsim2.5$ mag at H$\alpha$.
Another possibility is that at least some of the mid-infrared emission
in these sources arises from a nonthermal component \citep[e.g.][]{A1689}.

\begin{figure}
\leavevmode \epsfysize=8.5cm \epsfbox{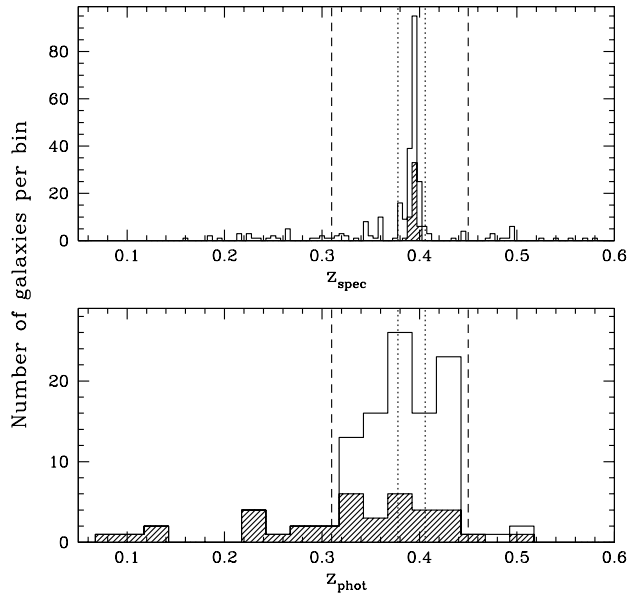}
\caption{
{\bf Bottom panel: }
The photometric redshift distribution for the spectroscopically
confirmed members is shown as the  open histogram.  The
spectroscopic cluster redshift limits ($0.378<z<0.406$) are shown as
the vertical, dotted lines.  The photometric redshift distribution is
broader; we take the photometric redshift cluster limits to be
$0.31<$\zphot$<0.45$, as indicated by the vertical, dashed lines.
The solid histogram shows the distribution for \halpha\ emitters.
{\bf Top panel:} 
The distribution of spectroscopic redshifts for galaxies with
$0.31<$\zphot$<0.45$, indicated by the  dashed lines, is shown as
the open histogram.  The solid histogram shows the distribution for
all \halpha\ emitters.
\label{fig:photz}
}
\end{figure}

\subsection{Photometric redshifts and field contamination}\label{sec-photz}

To determine cluster membership for all of the galaxies within our
imaging field of view 
we use  broad-band colours to compute photometric redshifts, \zphot, 
using the code of \citet{KBB}. 
The reliability of these redshifts can be tested by comparing
with the spectroscopic catalogue of \citet{Czoske_cat}.
Fig.~\ref{fig:photz} shows the distribution of \zphot\ for
spectroscopically confirmed cluster members ($0.378<z<0.406$).
We note that this redshift range for spectroscopic members is set so that
the transmission of our narrow-band filter is better than 8\% for
the H$\alpha$ lines from those galaxies.
This corresponds to the velocity range of
$-3400<\Delta{v}<2300$\,km\,s$^{-1}$
and therefore it includes a significant fraction of the substructures
reported in Czoske et al.\ (2001) (see our Fig.~1).
As seen in Fig.~\ref{fig:photz},
most of the members have $0.31<$\zphot$<0.45$; however, there is a small
tail of galaxies for which the
photometric redshift is $z<0.3$.  These are preferentially blue
galaxies, for which photometric redshifts are less accurate due to their 
flat spectra.

\begin{figure}
\leavevmode \epsfysize=8.5cm \epsfbox{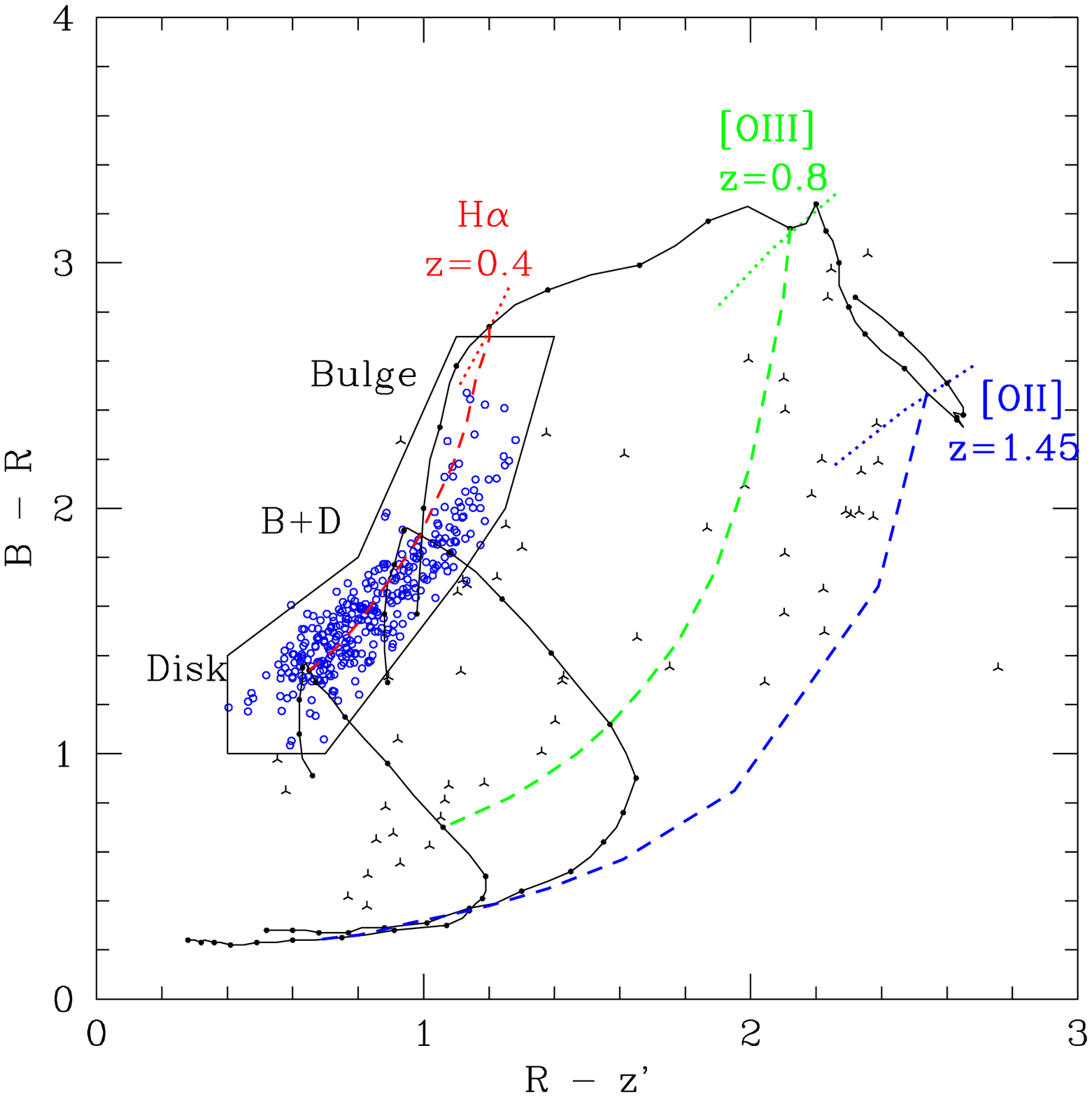}
\caption{
The colour--colour diagrams on which we can distinguish \halpha\ emission
of the cluster members from background \oii\ and \oiii\
contaminants.
Models of \citet{KBB}, as a function
of redshift, spectral type and metallicity, are shown by the lines.
Each contiguous solid line shows a model for a given spectral type,
at a series of redshifts between $0\le z\le2$.
The dashed lines connect the three spectral models
at fixed redshifts corresponding to H$\alpha$, [O{\sc ii}] and
[O{\sc iii}] emission, as indicated.
The dotted lines show the effect of changing metallicity
in the pure, passive (``Bulge'') model between $-$0.52$\le$[M/H]$\le$0.06.
The observed colour distribution of the strong emitters
($z^\prime-NB_{912}>0.25$) brighter than $z^\prime=21.8$ (fainter
than this limit the photometric errors increasingly dilute the visible
structure)
in the Cl\,0024 field are plotted over the model grids.
We identify the likely H$\alpha$
emitters {\it (open circles)} as either those
whose photometric redshifts are  within our membership
slice or those
whose colours are within the box drawn along the $z\sim 0.4$ sequence.
The latter galaxies are regarded as cluster members
even if they have apparently inconsistent photometric redshifts.
We also show the distribution of
narrow-band emitters which are not identified as cluster
members {\it (three-point crosses)} and are thus likely to be higher
redshift galaxies.
\label{fig:brz}}
\end{figure}

We want to ensure our sample of H$\alpha$ detections is not compromised
by incorrect photometric redshifts for blue
galaxies.  We can do this by noting that a detected emission line is most
likely to be either H$\alpha$ at $z\sim 0.4$, \oiii$\lambda5007$ at
$z\sim 0.8$ or \oii$\lambda3727$ at $z\sim1.5$.
Distinguishing these redshifts photometrically is
relatively easy, as shown in Fig.~\ref{fig:brz}.
The position of galaxies in the $(R-z^\prime)$--$(B-R)$ colour plane, as
a function of redshift, spectral type, and metallicity, are computed
using the models of \citet{KBB}.  
We also show the data, for all galaxies
detected in \nb912; these show a dominant sequence close to the model
prediction for \halpha\ emission, with distinct, secondary sequences
corresponding to the background galaxies with \oii\ or
\oiii\ emission.  We therefore consider all 511 galaxies that are detected
in the continuum--subtracted \nb912\ filter and have colours within the box
shown in Fig.~\ref{fig:brz} to be cluster members, whatever their 
photometric redshifts.  
We note that the higher redshift emission galaxies are distributed
uniformly over the field of view compared to the 
emission line members, confirming that they are unlikely to be
misidentified cluster members.
We can use these background galaxies to construct the \oii\ luminosity
function at $z\sim 1.5$, which we leave for future work. 

We will therefore define cluster members to be galaxies with 
$0.31<$\zphot$<0.45$, or those detected in H$\alpha$.
We also include all spectroscopic members, with redshift quality flags
A(secure) or B(probable), from the catalogue of \citet{Czoske_cat}.
This combined technique of using both broad-band and  narrow-band
photometric selection appears to
work well in picking out cluster members,
as demonstrated by Fig.~\ref{fig:photz}.
We have
a slight bias against 
blue member galaxies in our photometric redshifts derived purely from
the broad-band data; however this is compensated by
the fact that these galaxies tend to be strong emission line
sources, and so are recovered by including the narrow-band filter.
This combination
is potentially a very powerful tool for future applications to 
study the whole range of spectral types in higher
redshift structures using  optical/near-infrared filters.

We identify 2385 cluster members defined in this way down to $z^\prime=22.7$,
over 712.79~arcmin$^2$, for a mean projected density of
33 Mpc$^{-2}$.  Fig.~\ref{fig:photz} shows the distribution of
spectroscopic redshifts for these photometric cluster members.  
Most
of the galaxies are true cluster members ($0.378<z<0.406$), although
$\sim 30$ per cent are background and foreground galaxies that are
misidentified as cluster members.  
We correct for this residual
contamination based on the control field, the Subaru Deep Field
\citep{SDF1_short}.
The same photometric redshift estimator is applied
to galaxies in this field, observed in the same
combination of passbands to estimate the number of field galaxies
within the cluster redshift limits $0.31<$\zphot$<0.45$.  
We detect 1300 galaxies with magnitude $z^\prime<22.7$ in this redshift
range, over 667.94~arcmin$^2$.
This corresponds to a background object density of 19 Mpc$^{-2}$,
which is our contamination; this is
statistically subtracted from our photometrically--determined cluster
members, as a function of magnitude and colour,
as described in \citet{Kodama_cl0939}.

\begin{figure*}
\vspace{2cm}
\begin{center}
Fig.~8 is provided in jpg format.
\end{center}
\vspace{2cm}
\caption{
The projected distribution of the photometrically--determined cluster
members in Cl\,0024.
North is top and east is to the left.
Filled circles are strong emitters (\ewha$>40$~\AA) brighter than
$z^\prime$=21.8
and the {\it small dots} are the others.
The red and blue points separate the galaxy colours redder or bluer than
$(B-R)_c=2.0$.
The contours trace lines of constant projected galaxy density,
corresponding to 30, 50, 70, 100, 200 and 300 Mpc$^{-2}$ after residual
field contamination is subtracted.
The virial radius of this cluster (1.7~Mpc) is shown by the dotted circle.
\label{fig:map}}
\end{figure*}

\subsection{Local density measurements}\label{sec-denmeas}

The local surface number density 
of cluster members is calculated from the 10 nearest neighbors
(including that galaxy) brighter than our magnitude cut, $z^\prime=22.7$.  
We correct for residual field contamination in the redshift slice by
subtracting the mean density of galaxies in the control field, 19 Mpc$^{-3}$.
Unlike complete redshift surveys, we are not sensitive to densities
near or below the average field density, because the photometric redshift
slice is too wide to allow us to probe low physical densities.  

It will be interesting to compare our results with similar measurements
in the local Universe
\citep[e.g.][]{2dF_short,Sloan_sfr_short,2dfsdss}.  To compare
projected surface densities with these surveys, we need to account for
two fundamental differences in the way they are measured.  The first is
that, despite the background
subtraction, the measured density is still a projected quantity over a
large redshift slice, which corresponds to 855 Mpc (co-moving).
Local densities measured from redshift surveys, however, typically project
over only $\pm 1000\kms$, which corresponds to $\sim 30$ Mpc.  The
projected volume in the present survey, therefore, is $\sim 30$ times
larger than the corresponding volume in the low redshift surveys, and
we can expect our densities to be larger by a similar factor.
Note however that the actual factor is likely to be smaller than this,
since the galaxies would not be uniformly distributed within our redshift
slice; rather they are likely to be structured in discrete redshift
slices.
Secondly, the present survey extends to $M^\ast+3.5$,
which is
2.5 mag fainter (relative to $M^\ast$) than
the local Universe study of \citet{2dfsdss}.  Assuming a luminosity
function with faint end slope
$\geq -1.05$ \citep{Blanton03}, we expect $\gs 1.7$ times more galaxies in
our deeper survey.  Accounting for both these effects, we expect our
local projected densities to be $\gs 50$ times 
larger than those of \citet{2dfsdss}.

\section{Results}\label{sec-results}

\subsection{Spatial Distribution}

\begin{figure*}
\leavevmode \epsfysize=10.0cm \epsfbox{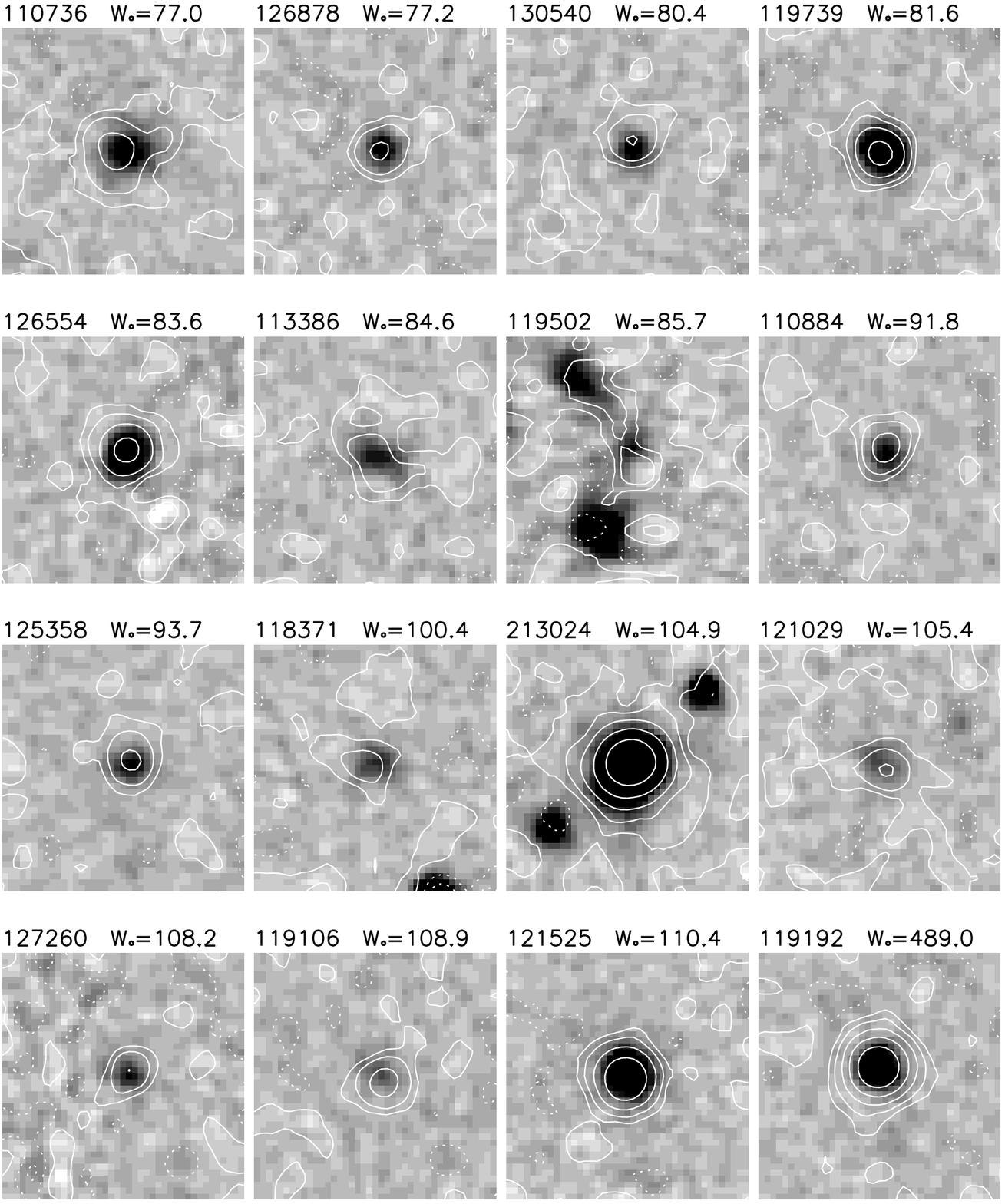}
\hskip 0.5cm \epsfysize=10.0cm \epsfbox{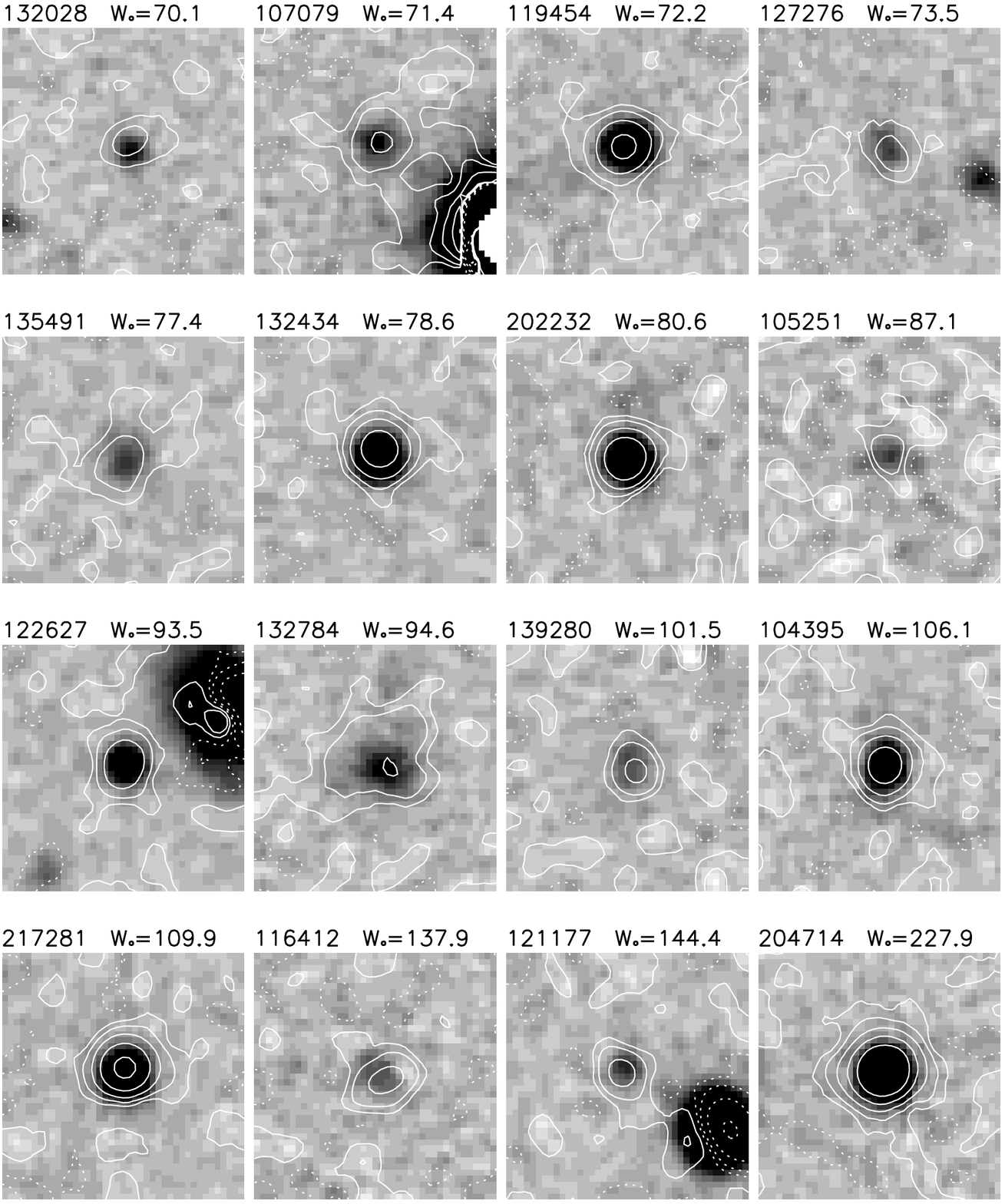}
\caption{{\it Left:} The greyscale images are $z^\prime$-band continuum
  images of galaxies within 1.7 Mpc ($R_{200}$) of the cluster centre
  with the strongest H$\alpha$ emission.  Images
  are 8~arcsec $\times$ 8~arcsec (42.7$\times$ 42.7 kpc).
  North is top and east is to the left.
  The {\it solid contours} are logarithmically
  spaced \halpha\ isophotes; the {\it dotted contours}
  represent H$\alpha$ in absorption.
  The ID number and \ewhanii\ of the galaxy are
  labelled at the top of each panel. {\it Right:} The same, but for
  galaxies within $1<R/R_{200}<2$.
\label{fig:im_cluster}}
\end{figure*}

The projected galaxy distribution of Cl\,0024 is shown in
Fig.~\ref{fig:map} for all cluster members.  Contours
of constant projected galaxy density are overlaid. 
The main cluster body, dominated by red galaxies, appears elongated in the
NW--SE direction.
On a larger scale, we see filamentary substructures towards the North--NW
and to the East, suggesting that the cluster is still dynamically evolving
by accreting  galaxies/groups from its immediate surroundings. This is
qualitatively similar to the
distribution of galaxies around A\,851 at $z=0.41$
(Kodama et al.\ 2001).
It is notable that the group to the NW of the cluster core shows a 
large fraction of galaxies detected in \halpha.

In Fig.~\ref{fig:im_cluster} we show images of the strongest
\halpha\ emitters within
$R_{200}$ (1.7 Mpc) and within the ``infall region'' $1<R/R_{200}<2$.
In most cases
we detect extended emission, suggesting that it does not exclusively arise
from AGN activity.  Furthermore, the H$\alpha$ emission generally traces the
optical structure of the galaxy, although there are some galaxies in
which the \halpha\ emission appears to be offset by a small amount
($\lesssim 2$ kpc) from the continuum light.

\subsection{{\it Hubble Space Telescope} Morphologies}

The available wide-field {\it HST} mosaic
imaging \citep{Treu03} affords
us the opportunity to compare the star
formation activity in Cl\,0024 with galaxy morphology.  Since
environmental effects are likely to influence morphology and star
formation on different time-scales, we might expect a change in the
correlation between the two properties, as a function of density
and/or redshift.
Unfortunately, we do not have equivalent morphological information for
our control field sample, so we cannot make the field correction
necessary to reliably determine the morphology
distribution as a function of cluster environment for the whole
sample.   For
galaxies that are undetected in H$\alpha$, we are therefore restricted to 
the spectroscopically-confirmed
cluster members \citep{Czoske_cat}. The population of galaxies that
are detected in H$\alpha$, however, does
not require such a correction, since we are able to identify
almost all background galaxies detected in emission (see
Fig.\ref{fig:photz}), and we can therefore use the full sample. 

\begin{figure}
\leavevmode \epsfysize=8.5cm \epsfbox{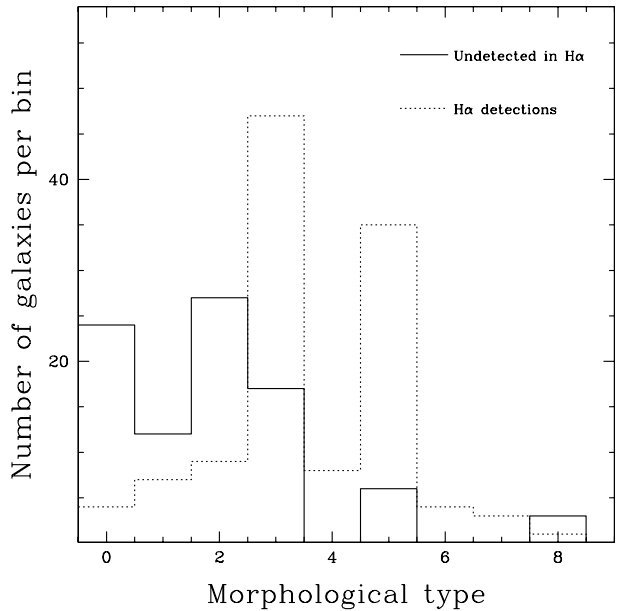}
\caption{The distribution of morphological classes, for
  spectroscopic members
  with no detected H$\alpha$ emission (solid line), and all galaxies
  with detected H$\alpha$  (dotted line).
The morphological
classes correspond to visual morphologies as follows:
0=E, 1=E/S0, 2=S0, 3=Sa+b, 4=S, 5=Sc+d, 6=Irr, 7=unclassified, and 8=merger,
respectively \citep{Treu03}.
\label{fig:morphSFR2}}
\end{figure}

Fig.~\ref{fig:morphSFR2} shows the  distribution of
morphological classes for the 
H$\alpha$ detections and non-detections \citep[using
the scheme of][]{Treu03}.  There is a distinct separation, as expected,
in that galaxies detected in H$\alpha$ tend to be late-type, while
undetected galaxies are early-type.
We do however see a substantial number of late-type, confirmed
cluster members without H$\alpha$ emission.  To investigate the
morphological properties of these galaxies in more detail we show
thumbnail images of 12 of the H$\alpha$-undetected members from the late
morphological classes ($\geq2$) in Fig.~\ref{fig:passiveS} and contrast
these with a random sample
of H$\alpha$-detected galaxies with the same distribution of types.
The first point to note is that the two H$\alpha$-undetected galaxies 
identified as mergers (T=8) both appear to contain 
early-type galaxies with very close companions. The absence of detectable
H$\alpha$ emission from these systems is not particularly surprising and
so we ignore them.  The morphological classifications for the bulk of the
mid/late-type H$\alpha$-undetected galaxies show no
strong differences from the matched sample of H$\alpha$-detected galaxies.
There is a suggestion that the H$\alpha$-undetected galaxies may be larger
systems (with 40 per cent of the sample having sizes in excess of 50\,kpc).
However, similarly large spirals are seen in the H$\alpha$-detected sample
and the small size of the undetected sample means this is not a
statistically compelling result. 
Another potential difference is  that none of the spiral galaxies
undetected in H$\alpha$ show much evidence for bright knots or other
clumpy structure associated with star--forming regions; the disks and
spiral arms appear smooth.  On the other hand, such features are
visible in some of the control galaxies.
We discuss the H$\alpha$-undetected
population further in \S~\ref{sec-passive}.

\begin{figure}
\vspace{2cm}
\begin{center}
Fig.~11 is provided in jpg format.
\end{center}
\vspace{2cm}
\caption{
The top three rows of the figure show thumbnail {\it HST} F814W images
of morphologically-classified mid/late-type and merging galaxies
which are spectroscopically-confirmed as cluster members, but which are
not detected in our H$\alpha$ narrow-band survey.  In the lower three
rows we compare these to a randomly-selected sample of H$\alpha$
narrow-band detected cluster members with an identical distribution of
morphological types.
The two mergers appear to be early-types with close companions (their
lack of H$\alpha$ emission is therefore unsurprising).  Otherwise, the
bulk of the undetected, mid-type members are morphologically
indistinguishable from the detected population, with the one caveat
that they appear to have larger physical scales (although the small
sample means this statement is not statistically significant). Each
panel is 10 arcsec square (53\,kpc) and is labelled with the morphological
class of the galaxy from \citet{Treu03} and either the redshift or
NB (for those members detected in the H$\alpha$ narrow-band image).
The panels are arbitrarily orientated and are presented at the raw
WFPC2 resolution, 0.17 arcsec, and default pixel scale,
0.1 arcsec\,pixel$^{-1}$.
\label{fig:passiveS}}
\end{figure}

\subsection{The \halpha\ Luminosity function}

Our sample of galaxies with \halpha\ emission is limited by apparent
continuum magnitude and equivalent width, as shown in
Fig.~\ref{fig:znb}.  From that figure, it is evident that our sample is
approximately complete for 
\lha$>1.3\times10^{41}$ ergs~s$^{-1}$, which corresponds to SFR $\gtrsim2$
$M_\odot$~yr$^{-1}$. Brighter than this limit, we
effectively have a volume-limited sample of \halpha--selected
sample of galaxies, and it is
straightforward to construct the luminosity function for Cl\,0024.  
Furthermore, most of the galaxies with luminosities brighter than this
limit have $z^\prime<21$, $1.7$ magnitudes
brighter than our survey limit, so the effect of photometric errors
are small.
At fainter line luminosities,
we begin to lose bright galaxies with low \ewhanii\ from the
sample; however, the luminosity function is dominated by fainter
galaxies with
higher EW and, hence, the effect is small.
Below \lha$\sim6\times10^{40}$ ergs~s$^{-1}$, our incompleteness increases
further because we do not observe galaxies with $z^\prime>22.7$.  This
incompleteness starts to become severe below
\lha$\sim2\times10^{40}$ ergs~s$^{-1}$.  Photometric errors also start
to become important at this luminosity, because most of the
contribution is from galaxies near our magnitude limit.
We make no correction for
these effects, but focus on the shape of the luminosity function
brighter than \lha$\sim10^{41}$ ergs~s$^{-1}$.  

\begin{figure*}
\leavevmode \epsfxsize=12cm \epsfysize=12cm \epsfbox{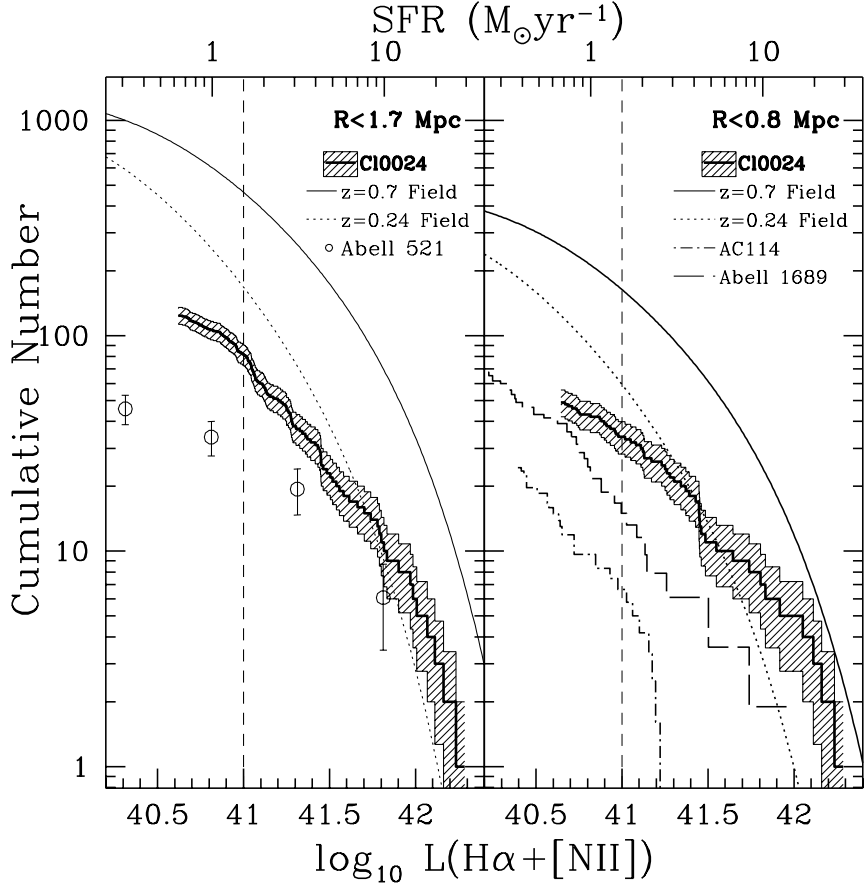}
\caption{The \halpha\ luminosity function in Cl\,0024 and 1$\sigma$
uncertainty {(shaded region)} within $R_{200}$ (1.7 Mpc, {left panel})
and 0.47$R_{200}$ (0.8 Mpc, { right panel}) of the cluster centre,
uncorrected for dust extinction and \nii\ contamination.
The {vertical, dashed lines}  show the approximate completeness limit
of our survey; fainter than this limit we underestimate the luminosity
function.  These results are compared with the Schechter function fits
to the field luminosity functions at $z=0.24$
\citep{Fujita_half} and $z\sim 0.7$ \citep{TMLC}, renormalised to the
theoretical overdensity of Cl\,0024 within the appropriate radius as
described in the text.  We also compare with the clusters A\,521
\citep[$z$=0.25,][]{Umeda03}, AC114 \citep[$z$=0.31,][]{C+01} and
A\,1689 \citep[$z$=0.18,][]{A1689}, as indicated in the legends.
\label{fig:half}}
\end{figure*}

The luminosity function of galaxies within $R_{200}=1.7$ Mpc (5.2~arcmin)
of the cluster centre is shown in the left panel of Fig.~\ref{fig:half}.  
It is well fit, for \lha$>10^{41}$ ergs~s$^{-1}$,  by a Schechter
function with \lshanii$=1.1\times10^{42}$ ergs~s$^{-1}$ and $\alpha=-1.5$.
Recall that this is uncorrected for dust extinction and \nii\ emission.
Making an approximate correction for 1 magnitude of dust extinction and
30 per cent \nii\ contribution gives \lsha$=2.1\times10^{42}$ ergs~s$^{-1}$.
The uncertainties on $\alpha$ and \lshanii\ are strongly correlated and
depend upon the magnitude range of the fit; thus we do not present our
parameters as an optimal choice, but merely as a way to compare the shape
with previous work.

We can compare this luminosity function with other clusters, at lower
redshift.  The cluster A\,521, at $z=0.25$, was studied with a very
similar technique, by \citet{Umeda03}.  We adjust their published
luminosity function to account for our different
cosmology, and to remove their corrections for \nii\ and dust
extinction.  We also apply a small rescaling to account for the
difference in physical area considered, as \citet{Umeda03} use a 2 Mpc
radius, under a different cosmological model.  The result is
shown in the left panel of Fig.~\ref{fig:half}; the characteristic
luminosity in their Schechter function fit with the above adjustments is
\lshanii$=3.9\times10^{41}$ ergs~s$^{-1}$, a factor three fainter than in
Cl\,0024.  We also compare with the luminosity functions of two other 
clusters, AC\,114 \citep[$z=0.31$,][]{C+01} and A\,1689
\citep[$z=0.18$,][]{A1689}.
These are based on deep spectroscopic observations made with the LDSS++
spectrograph on the Anglo-Australian Telescope, over a smaller field of
view (7~arcmin, corresponding to a diameter of 1.3 Mpc in
A\,1689, and 1.9 Mpc in AC\,114).  To make a fair comparison, we show
the luminosity functions within a radius of 0.80 Mpc, in the right panel of
Fig.~\ref{fig:half}.  The published AC\,114 and
A\,1689 luminosity functions are multiplied by a small factor 
to approximately correct to a circular area of 0.80 
Mpc radius.
We also apply a correction for aperture bias by comparing the total
$I-$band magnitudes to the magnitude within the spectroscopic aperture
size.  This amounts to a correction factor of $\sim 1.4$.  These luminosity
functions are steeper than that of A\,521, although with a similar
characteristic luminosity.  
All three of these lower redshift clusters
show $\gtrsim 4$ times fewer H$\alpha$ emitters of fixed luminosity
than Cl\,0024, within the same physical area.
However, it is not clear whether this is due to a
difference in cluster richness, or to a real evolutionary effect.
In \S~\ref{sec-discuss} we will show the evolution of cluster SFR
normalised to total cluster mass, to remove this degeneracy.

Finally, we compare our data with the field luminosity function at
$z=0.24$ and at $z\sim 0.7$.  The lower redshift data
are also obtained from
a narrow-band {Suprime-Cam} survey \citep{Fujita_half},
and are in good agreement with the
spectroscopically determined luminosity function at similar redshift
\citep{TM98}.  The higher redshift luminosity function is obtained from
spectroscopic measurements at the VLT \citep{TMLC}.
After making the usual
adjustments to remove the corrections for reddening and \nii\ contribution,
we renormalise to the expected average cluster density within $R_{200}$
(left panel) and $0.47R_{200}$ (right panel).  In the first case,
the average cluster density is $200\rho_{\rm crit}$, or $\sim 670$ times
the average field density (for $\Omega_m=0.3$), by definition.
For the smaller radius, we use the mass model of \citet{Kneib03}
to determine that the cluster overdensity
is $\sim 2260$ times larger than the average field density.
The $z=0.24$ field H$\alpha$ luminosity
function has a similar shape to the A1689 and AC114 clusters, with
\lshanii$=4.8\times10^{41}$ ergs~s$^{-1}$. On the other hand, the $z\sim
0.7$ function is brighter, with \lshanii$=6.3\times10^{41}$ ergs~s$^{-1}$,
and provides a close match to the Cl\,0024 data.  In all cases, the
renormalised field has $\gtrsim 5$ times more emitters of a given
luminosity than clusters at similar redshifts.
Therefore, we find that the shape of the H$\alpha$ luminosity function
does not depend strongly on environment, although the normalisation
does.  There is significant
evolution between $z\sim 0.2$ and $\sim 0.4$, corresponding to a factor
$\sim 3$ in H$\alpha$ luminosity.  This degree of evolution is seen in
both the field and cluster luminosity functions.

\subsection{Trends with local density and cluster-centric radius}
\label{sec-density}

\begin{figure}
\leavevmode \epsfysize=8.5cm \epsfbox{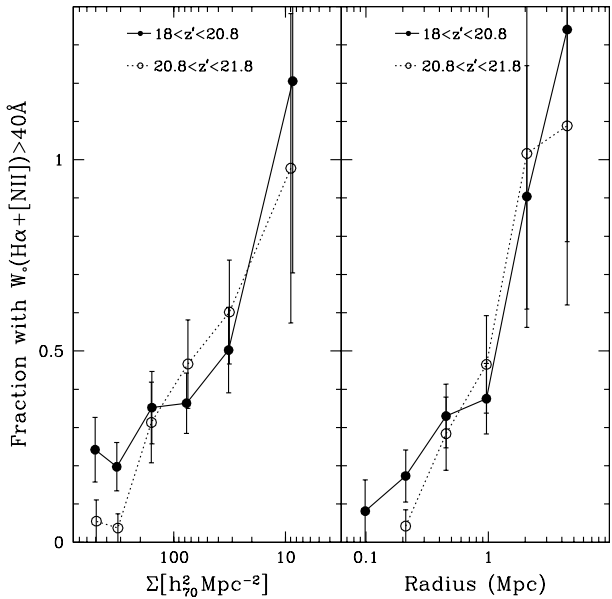}
\caption{The fraction of galaxies with \ewhanii$>40$~\AA, corrected for
residual field contamination,
is shown as a function of local projected
density (left panel), and cluster-centric radius (right panel),
for bright and faint galaxies separately.
\label{fig:fha}}
\end{figure}

We will now consider how the H$\alpha$ distribution depends on the environment
around Cl\,0024.  In Fig.~\ref{fig:fha} we show the fraction of galaxies with
\ewhanii$>40$~\AA\ as a function of local
projected density and cluster-centric radius.  We only consider galaxies
brighter than  $z^\prime=21.8$ because, fainter than this limit,
the photometric errors scatter a significant number of galaxies with
low equivalent widths to \ewhanii$>40$~\AA\ (see Fig.~\ref{fig:znb}).  
Both the density and the strong emission line
fraction are corrected for residual field contamination using the SDF control
field.  As expected, there is a decrease in the fraction of
strong emission line galaxies with  increasing density, toward the
cluster centre.  The same  trend is seen for both bright and faint
galaxies.  At low densities, $\Sigma\lesssim20$~Mpc$^{-2}$, the
statistical background correction dominates, and the uncertainties
become large
\footnote{Note that the statistical nature of the background subtraction
means the fraction of galaxies with H$\alpha$ emission can exceed unity
at the lowest density end.}.

\begin{figure}
\leavevmode \ 
\epsfysize=8.5cm \epsfbox{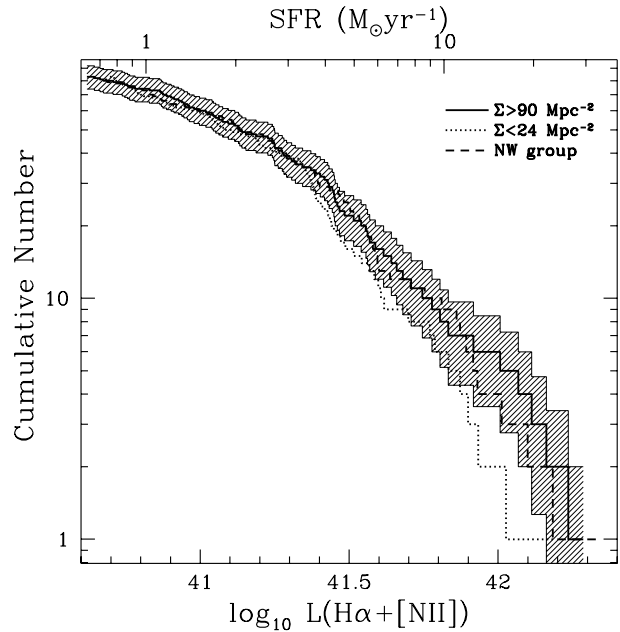}
\caption{The H$\alpha$ luminosity function 
in the highest--density regions ($\Sigma>90$ Mpc$^{-2}$, but excluding the
NW group), lowest--density regions ($\Sigma<24$ Mpc$^{-2}$),
and in the NW group of galaxies.
Each sample has 83 galaxies. The {\it shaded region}
represents the 1$-\sigma$ uncertainty on the luminosity function for
the dense region, only.  
\label{fig:harad}}
\end{figure}

Thus, the fraction of strong emission line galaxies depends strongly on
local environment.  
However, for those galaxies that {\it are} detected in emission, 
the equivalent width distribution and H$\alpha$ luminosity
function are approximately 
independent of environment.  This is shown
in Fig.~\ref{fig:harad}, where we compare the 
H$\alpha$ luminosity functions for
galaxies detected in H$\alpha$ within the highest density environments
(excluding the NW group), with
those  at the lowest densities. We also show the data for
galaxies in the NW group, which has an unusually high fraction of
emission line galaxies, consistent with 100 per cent
for blue galaxies
(see below). 
The H$\alpha$ luminosity function shows  little
sensitivity to local density; although the cluster and group appear
brighter by $\sim 0.2$ mag than the lowest density regions, this
is only a $\sim 1\sigma$ difference.
We conclude that emission line strength itself
does not depend strongly on environment, and 
it is only the fraction of emitters that changes with density.  The
same conclusion was reached from observations of the local Universe, by
\citet{2dfsdss}.

\begin{figure}
\epsfysize=8.5cm \epsfbox{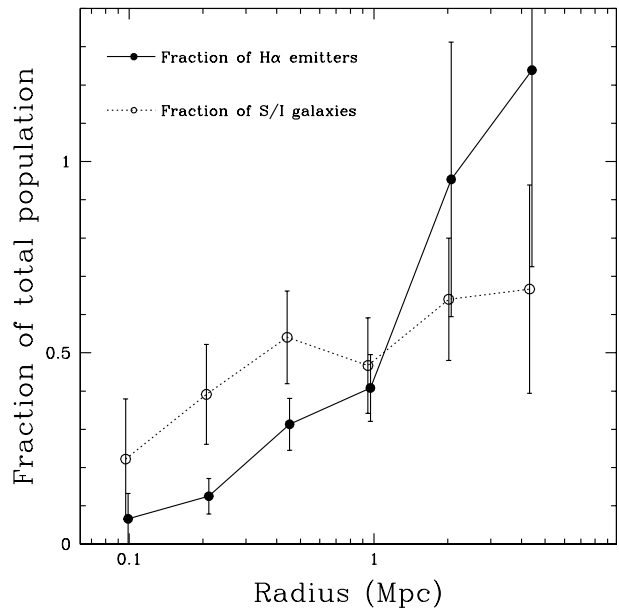}
\caption{The {\it solid circles} show the fraction of galaxies
  detected in \ha, statistically corrected for field contamination, as
  a function of cluster-centric radius.  The {\it open symbols} show the
  fraction of spectroscopically--confirmed cluster members classified
  as spiral or irregular types by \citet{Treu03}; these points are
  offset in radius by a small amount for clarity.  Both samples are
  restricted to magnitudes $z^\prime<21.8$.
\label{fig-sfrT}}
\end{figure}

In Fig.~\ref{fig-sfrT} we compare the fraction of spiral and irregular
galaxies (spectroscopic cluster members identified from {\it HST} images)
with the field--corrected fraction
of galaxies with significant \ha\ emission, as a function of
cluster-centric radius.  Both samples are limited to $z^\prime<21.8$.
As shown by \citet{Treu03}, the fraction of
spiral and irregular galaxies increases sharply within the inner $\sim
0.5$ Mpc, and remains fairly constant at $\sim 50$--$60$ per cent
beyond that radius.  In contrast with this, the fraction of galaxies
detected in \ha\ rises gradually with increasing radius, out to
$\sim 1$--$2$ Mpc, where it quickly increases to $\gtrsim 80$ per
cent.  This demonstrates that the trends in morphology and star
formation are at least partly independent, and that the underlying
mechanisms driving these trends may be different, as we discuss further
in \S~\ref{sec-passive}.

\begin{figure}
\epsfysize=8.5cm \epsfbox{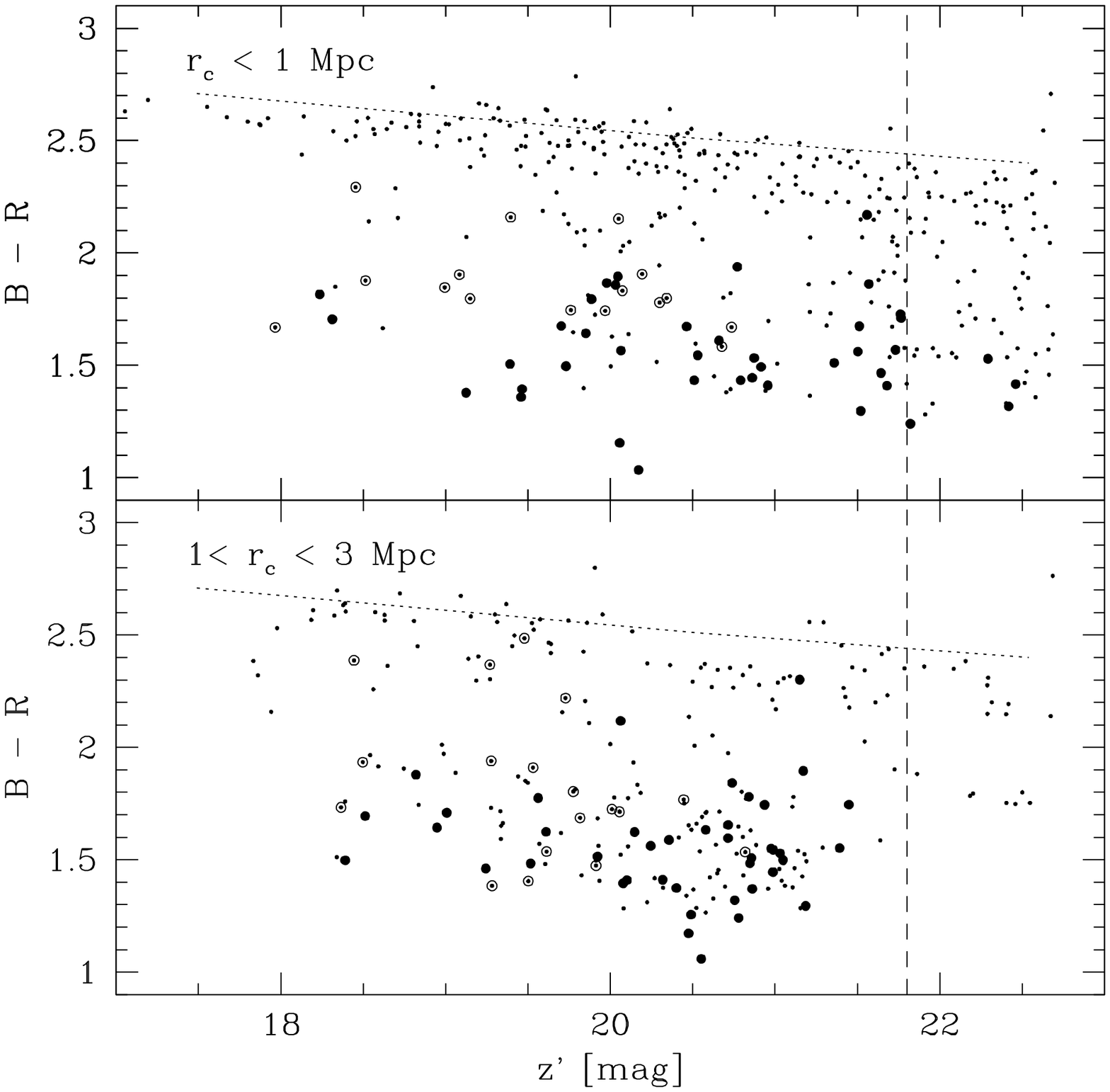}
\caption{The field-corrected colour--magnitude diagrams of galaxies in the
cluster core ({\it top panel}, $r_c<1$Mpc) and in the outskirts
({\it bottom panel}, $1<r_c<3$Mpc). The residual contamination is removed.
Galaxies detected with \ewhanii$\geq40$~\AA\ are shown
as {\it filled circles}; weaker emission lines, with \ewhanii$<40$~\AA,
are shown as {\it circled points} and undetected galaxies
are just points.
The dotted line illustrates the colour--magnitude slopes of the Coma
elliptical models \citep{KA97}.
The vertical dashed line indicates $z^\prime$=21.8 below which the strong
emitters become incomplete.
\label{fig:cmr}}
\end{figure}

\begin{figure*}
\leavevmode \epsfysize=8.5cm \epsfbox{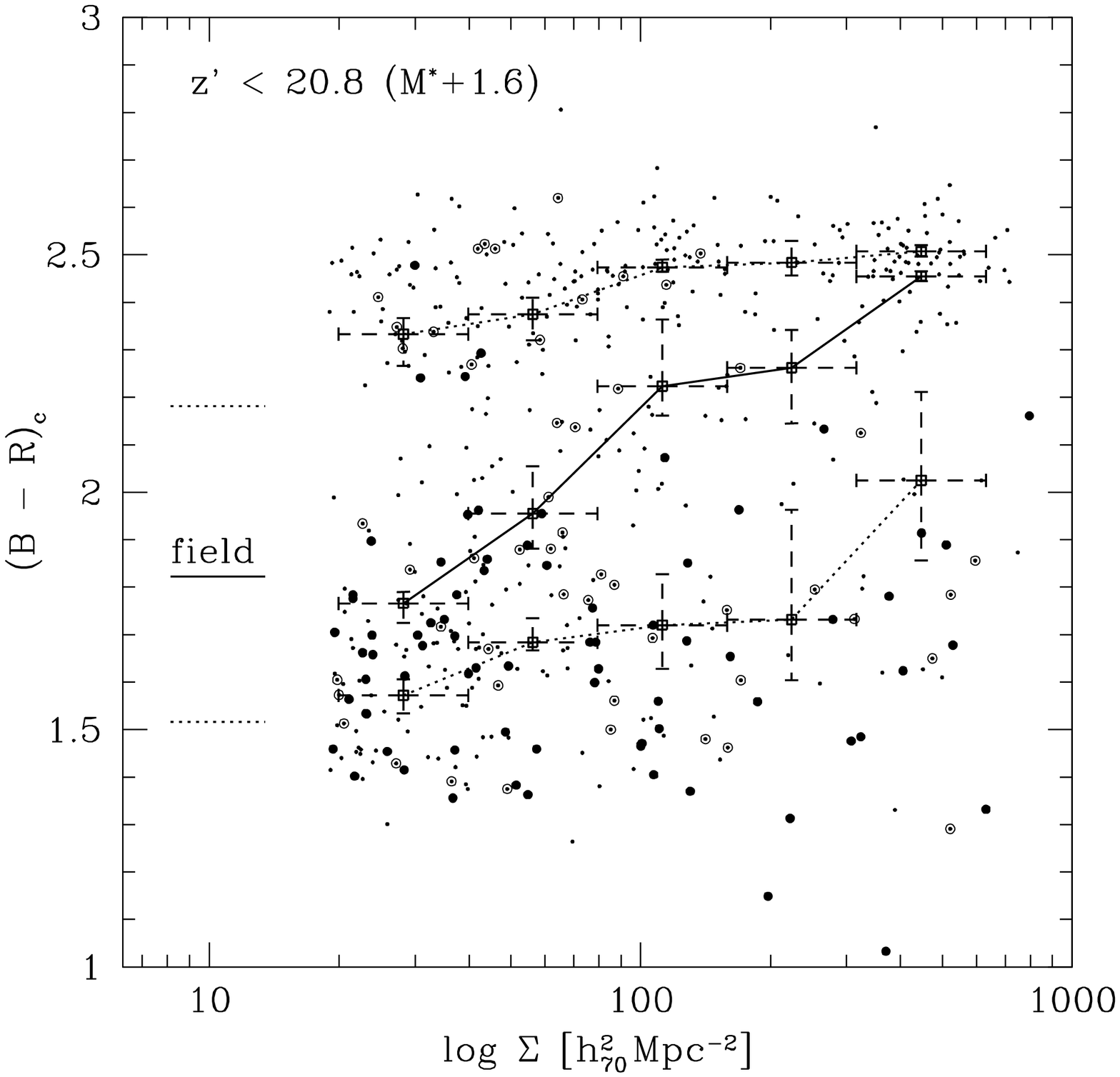}
\hskip 0.0cm \epsfysize=8.5cm \epsfbox{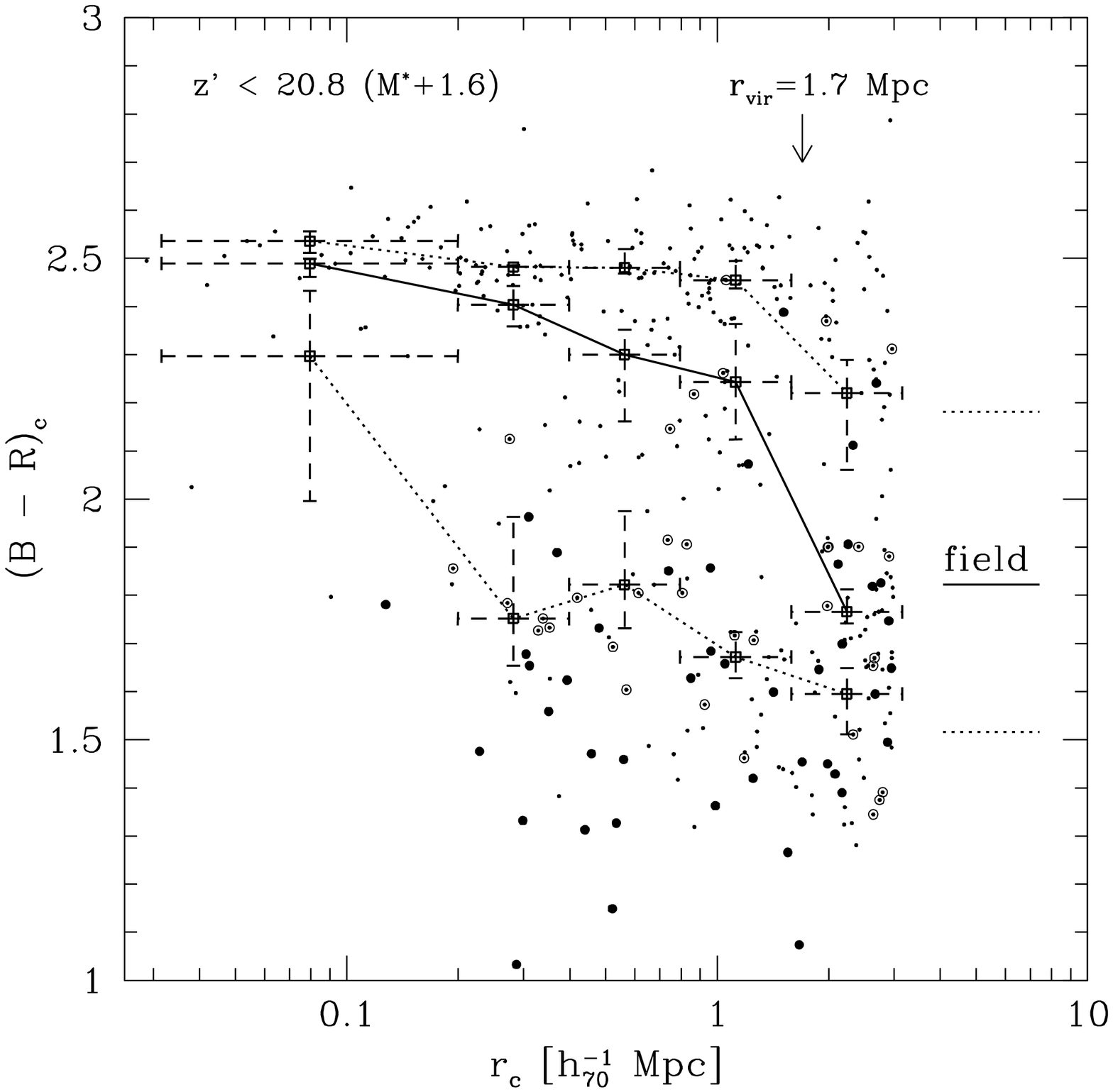}\\
\leavevmode \epsfysize=8.5cm \epsfbox{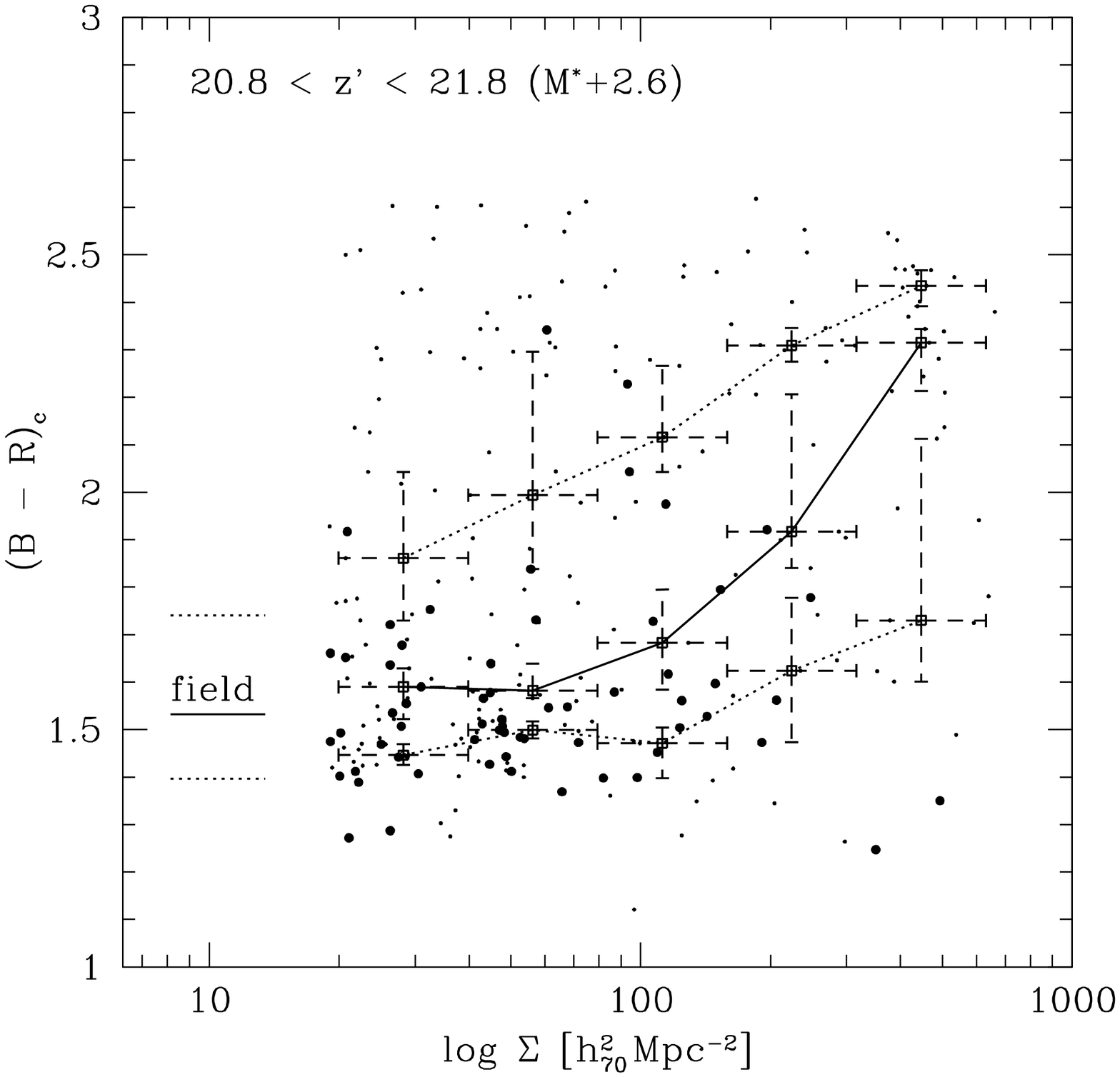}
\hskip 0.0cm \epsfysize=8.5cm \epsfbox{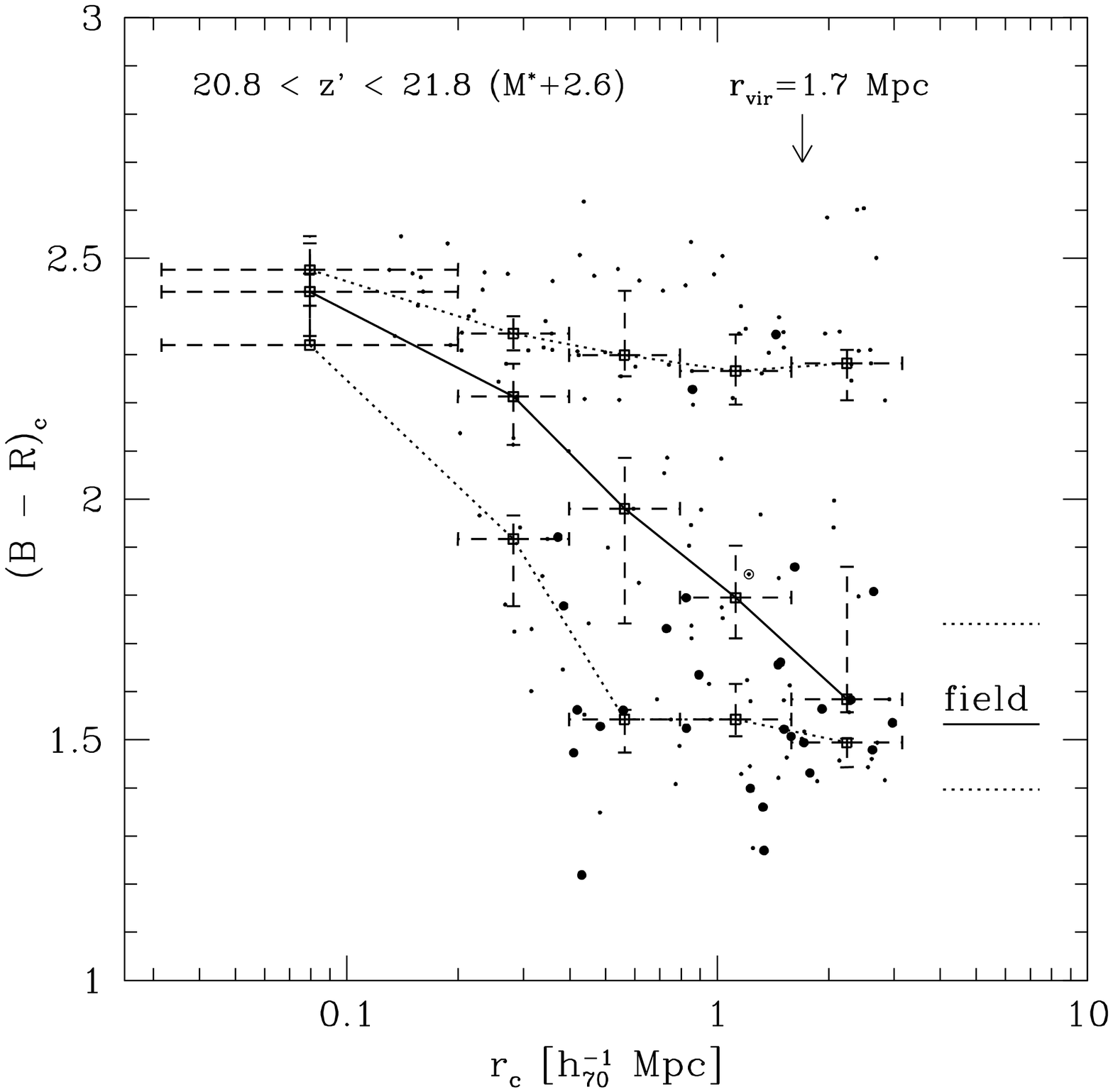}
\caption{The background corrected colour distributions $(B-R)_c$ of galaxies
as a function of local projected density ({\it left panels}) and
cluster-centric radius ({\it right panels}) are shown for bright galaxies
($M<M^\ast+1.6$, {\it top panels}) and faint galaxies
($M^\ast+1.6<M<M^\ast+2.6$, {\it bottom panels}).
The local density $\Sigma$ is calculated for 10 nearest neighbours
brighter than $z^\prime=22.7 (M^*+3.5)$ and the residual contamination is
removed.
Galaxies detected with \ewhanii$\geq40$~\AA\ are shown
as {\it filled circles}; weaker emission lines, with \ewhanii$<40$~\AA,
are shown as {\it circled points} and undetected galaxies
are just points.
The three folded lines denote the 25th ({\it upper dotted line}),
50th ({\it solid line}) and 75th ({\it lower dotted line}) percentile colour
loci with poissonian error-bars.
The three horizontal lines at the edges
show the percentile colours of galaxies in the SDF control
field with $0.31<$\zphot$<0.45$, the same cut as used for the cluster.
\label{fig:br_env}}
\end{figure*}

The integrated broad-band colours of the galaxies trace star-formation
on longer time-scales ($\gtrsim 0.5$ Gyr, although this is
model and waveband dependent) than H$\alpha$, and thus provide a useful
comparison.
Figure~\ref{fig:cmr} shows the colour--magnitude diagrams for galaxies
in two different environments.
As observed in the local Universe \citep[e.g.][]{Baldry03,BB04}
and at higher redshift \citep[e.g.][]{BellGEMS}
the galaxies form two distinct populations:
a blue population with $B-R\sim1.6$ (mostly showing emission)
and a red population with $B-R\sim2.4$ (having no or little emission),
and a gap is seen in between.
The relative proportion of these two populations
depends strongly on environment.  In the cluster core, the red
colour--magnitude sequence is much more populated and well established
than that seen in the outskirts, and the fraction of blue star forming
galaxies is much lower.
To see this environmental dependence more clearly and quantitatively,
we plot the $(B-R)$ colour as functions of local density and
cluster-centric distance in Fig.~\ref{fig:br_env}.
To concentrate on the environmental dependence of colours, we correct
for the slope of colour--magnitude relation (the dotted line on
Fig.~\ref{fig:cmr}) on the red sequence
using the models of \citet{KA97},
denoted $(B-R)_c$, where \begin{equation}
(B-R)_c = (B-R) + 0.043 (z^\prime-20.2),
\end{equation}
corresponding to the colour at $M^*+1$ ($z^\prime=20.2$).
The residual field contamination is corrected on these diagrams
using the SDF control field data in the same manner as adopted in
\citet{Kodama_cl0939}.
The densities on the horizontal axis are also corrected for the contamination.

Again, the blue population dominates at low densities/outer regions,
while the red population dominates at high densities/inner regions;
in fact, within $\sim200$ kpc of the centre, there are almost no blue galaxies.
It is clearly seen that the colour distribution of galaxies starts to become
predominantly blue 
at low densities or at large radii from the cluster centre, as also
seen in A\,851 at $z=0.41$ \citep{Kodama_cl0939}.
In fact, a transition from a predominantly blue population to one
dominated by red galaxies is identified at around
the virial radius (1.7~Mpc).  Although a strong sequence of red galaxies
is present in all environments, the blue galaxies quickly disappear within
this radius, where the relative number of red galaxies is much larger than
in the field populations (SDF), shown by the horizontal lines at the edge
of the diagrams.  This traces the trend observed in H$\alpha$ emission,
shown in Fig.~\ref{fig-sfrT}.

The galaxies with detected H$\alpha$ are preferentially blue; this is
particularly true for the fainter population, where almost all the
strong emission line galaxies are bluer than $(B-R)_c=2$.  On the other
hand, the brighter galaxies show a significant population of red
galaxies with detected emission.  Almost all of these have only
relatively weak emission, with \ewha$<40$~\AA, which would not be
detectable in the fainter galaxies.

\begin{figure}
\leavevmode \epsfysize=8.5cm \epsfbox{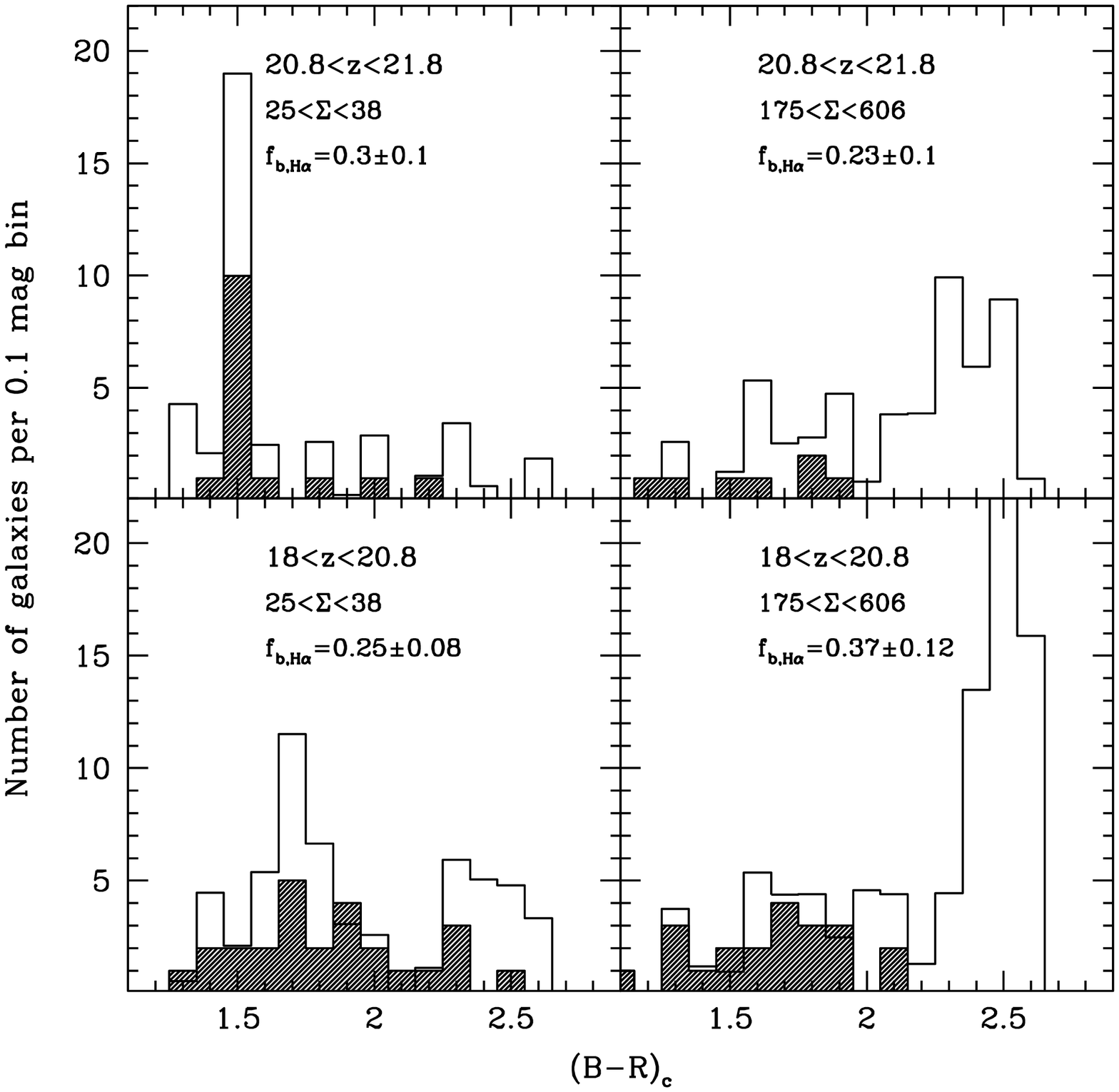}
\caption{The $(B-R)_c$ colour distribution of Cl\,0024, as a function of
luminosity and local, projected density ($\Sigma$, in units of
Mpc$^{-2}$), is shown as the {\it open histograms}.  The {\it solid
histograms} show the subset of galaxies detected in H$\alpha$.
In each panel, we label the value of
$f_{b,H\alpha}$, which is the fraction of galaxies bluer than
$(B-R)_c=2$ with \ewha$>40$~\AA.
\label{fig:coldist}}
\end{figure}

\begin{figure}
\leavevmode \epsfysize=8.5cm \epsfbox{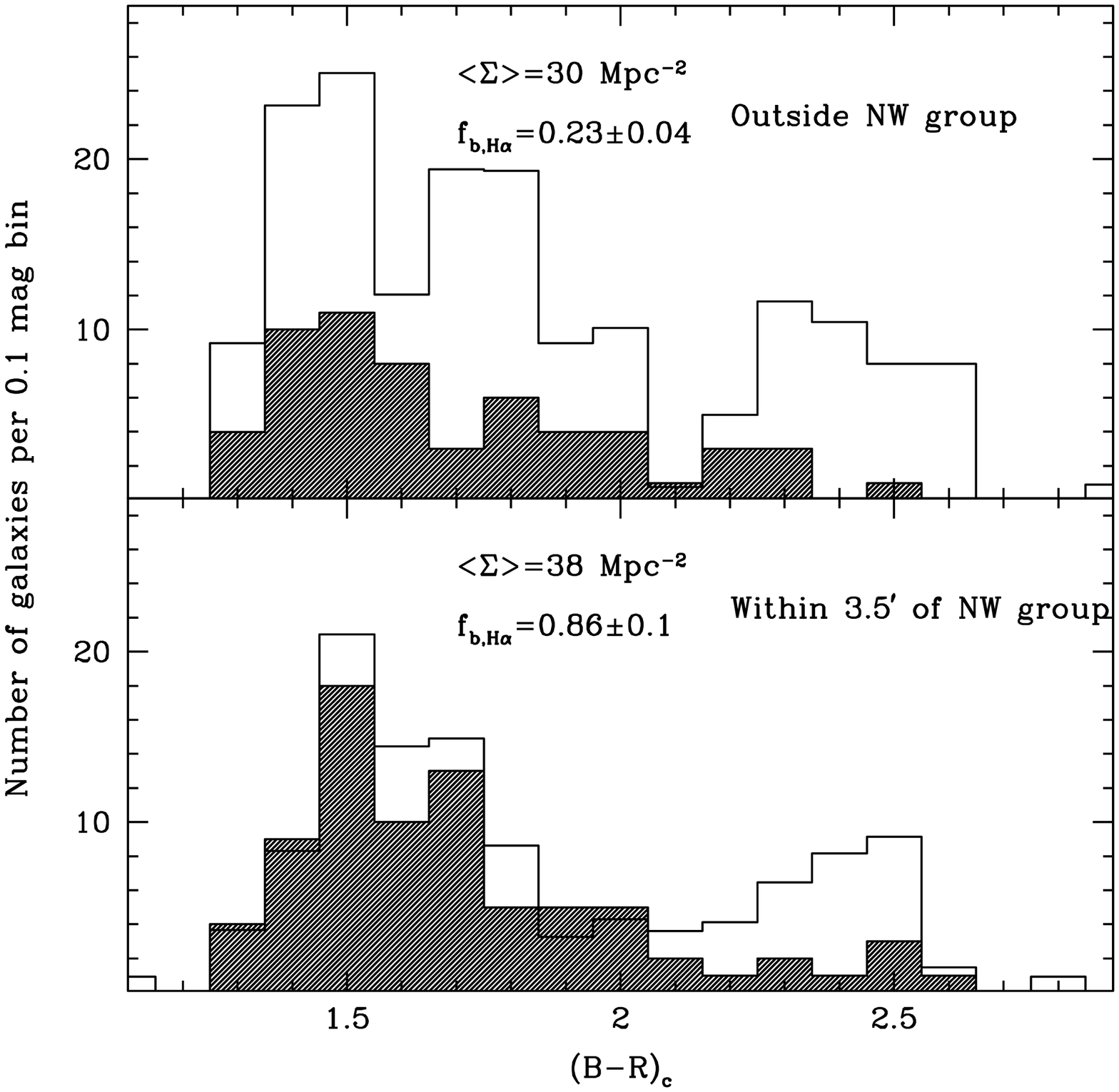}
\caption{The background--corrected colour distribution of galaxies in
the NW clump ({\it bottom panel}) and in regions outside this clump,
but at similar local projected densities ({\it top panel}).  The {\it
shaded histograms} show the colour distribution of galaxies
detected in H$\alpha$ emission.  In each panel we label the average
value of $\Sigma$, and the fraction of blue galaxies ($(B-R_c)<2$)
with H$\alpha$ emission.
\label{fig:clump}}
\end{figure}

These correlations are shown also in Fig.~\ref{fig:coldist},
where we plot the colour distributions as a function of luminosity and
environment.
The bimodality in the colours is clear again here,
as there are few galaxies with intermediate colours, $(B-R)_c\sim2$ at
any magnitude or in any environment.
In a fixed magnitude range, the relative height of the two distributions is
a strong function of density. 
Fig.~\ref{fig:coldist} also shows the colour distribution of
galaxies detected in H$\alpha$ emission;
as expected, they are almost
all blue galaxies.  We have computed the fraction
of blue galaxies [$(B-R)_c<2$] with \ewha$>40$~\AA, corrected for
residual field contamination, in each of our luminosity and environment
bins.  Approximately $30\pm10$ per cent of blue galaxies show strong
H$\alpha$ emission with no strong dependence on luminosity or environment.
The group of galaxies to the NW is therefore notable,
in that it has a very high fraction of blue galaxies.
In Fig.~\ref{fig:clump} we show the colour distribution of galaxies within
3.5~arcmin (1.1 Mpc) of this clump, corrected for residual background
contamination.  The population is statistically consistent with almost
all the blue galaxies having strong H$\alpha$ emission, \ewha$>40$~\AA.
We compare this with
the galaxy population outside of this clump, but with similar local
projected densities, $\Sigma=22.8$--$40$ Mpc$^{-2}$.
Here, only 23$\pm 4$ per cent of the blue galaxies show \ewha$>40$~\AA.
We note however that this difference can be due to the case that the NW
clump has just right systemic velocity in the line of site so that the
\halpha\ emission is neatly covered by the \nb912 filter, whereas the
regions outside of this NW clump may contain systems with different
velocities so that \halpha\ detection is incomplete (Fig.~\ref{fig:velocity}).

\section{Discussion}\label{sec-discuss}

\subsection{Evolution}

The dependence of H$\alpha$ emission on environment in the local
Universe has recently been quantified over a large range of scales,
using the 2dF galaxy redshift survey and the Sloan Digital Sky Survey (SDSS)
\citep{2dF_short,Sloan_sfr_short,2dfsdss}.   In particular,
\citet{2dfsdss} have shown correlations of galaxy population with local
projected density, $\Sigma_5$, which is computed from the distance to
the fifth--nearest (projected) neighbour within $1000\kms$.  They
showed that the fraction of emission line galaxies
decreases from $\sim 60$ per cent in low-density environments
($\Sigma_5=0.1$ Mpc$^{-2}$) to $\sim 10$ per cent in the
highest-density regions ($\Sigma_5=10$ Mpc$^{-2}$).  As discussed in
\S~\ref{sec-denmeas}, our densities
can be up to a factor of $\sim 50$ larger than theirs,
because of the larger projected volume and the deeper photometry.
With this correction (i.e.\ a factor of 50),
our observed correlation between
the fraction of galaxies with H$\alpha$ emission and $\Sigma$
(Fig.~\ref{fig:fha}) is in
good qualitative agreement with the local relation.  
A more precise
comparison is difficult because, in addition to the uncertain
correction to the density scale, \citet{2dfsdss} measure the
fraction of galaxies with \ewha$>4$~\AA, while our photometric
uncertainties limit our analysis to the fraction with much stronger
emission, \ewhanii$>40$~\AA\ (e.g. Fig.~\ref{fig:fha}).
Additional uncertainties
are associated with the selection band and the magnitude range of the
surveys.  However, it appears that the correlation between galaxy
population and local density has not evolved radically between $z\sim0.4$
and $z\sim0.1$.  

\begin{figure}
\leavevmode \epsfysize=8.5cm \epsfbox{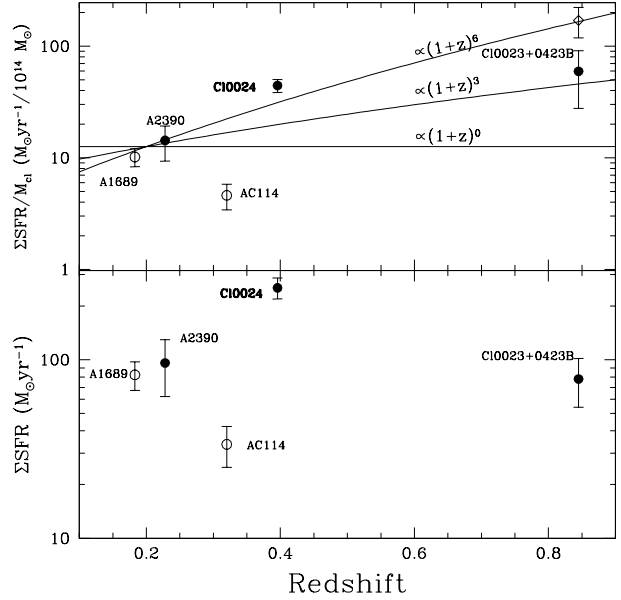}
\caption{{\bf Bottom panel:} The total SFR within 0.5 $R_{200}$,
is shown as a function of redshift for Cl\,0024 and four other clusters
with deep \halpha\ observations,
either from imaging data ({solid circles}) or aperture
spectroscopy ({open circles}). {\bf Top panel:} We show the total
SFR within 0.5$R_{200}$ divided by the total mass $M_{200}$, for the
same clusters.   The uncertainties are based on the number of
galaxies contributing to the sum, and do not include the
uncertainties on the mass estimates.  The {diamond symbol}  shows
the effect of assuming the mass estimate of \citet{FZM} for
CLJ0023+0423B, rather than that of \citet{PLO-II}.
The {solid lines} shows different scalings with $(1+z)$, as a guide.
\label{fig:FZM}
}
\end{figure}

Recently, \citet{FZM} attempted to compare the 
total SFR within clusters as a function of redshift and velocity
dispersion.  To make a similar comparison, 
we measure the total amount of star formation within 0.5$R_{200}$,
which corresponds to $\sim 0.85$ Mpc for Cl\,0024.  Within this
radius, we detect 56 galaxies and a total SFR of $\sim 253\pm 34$
M$_\odot$~yr$^{-1}$.  Most of this SFR ($>97$ per cent) comes from galaxies
with SFR$>1$ M$_\odot$~yr$^{-1}$, where our sample is complete.
Furthermore, even including all galaxies with \ewhanii$>0$~\AA\ in the
total, a clear overestimate due to large photometric errors at faint
magnitudes, only increases the total SFR by $\sim 17$ per cent.
Therefore, we do not expect our
incompleteness to have a large effect on this calculation.  

Following \citet{FZM} we compare this number
with  the total SFR within $0.5R_{200}$ of the clusters
A\,2390 \citep{A2390_BM}, A\,1689 \citep{A1689}, AC\,114 \citep{C+01}, and
Cl\,J0023+0423B \citep{FZM}.  In so doing, we include a correction for
sampling fraction and aperture bias for the spectroscopic results of
A\,1689 and AC\,114, that was neglected by \citet{FZM}; this amounts to an
increase in the total SFR by a factor $\sim 2.8$ in these clusters.
We adopt the $R_{200}$ given by \citet{FZM}, except for A\,1689, where we
take $R_{200}=1.6$ Mpc from \citet{KCS}.

In the bottom panel of Fig.~\ref{fig:FZM}, we first compare the total
SFR within 0.5$R_{200}$ of these five clusters.  There is no apparent
trend with redshift.  There is a large scatter -- 
with AC\,114 and Cl\,0024 differing by nearly an order of magnitude 
between the two clusters. 
We now follow \citet{FZM} and normalise these SFR totals by the cluster
masses.  For Cl\,0024, A\,1689 and AC\,114 we will use the lensing masses of
5.7$\times 10^{14}M_\odot$ \citep{Kneib03}, 8.1$\times 10^{14}M_\odot$
\citep{KCS}, and 7.3$\times 10^{14}M_\odot$ \citep{NKSE}, respectively.
For A\,2390 we use the X-ray mass derived from {\it Chandra} observations,
13.6$\times 10^{14}M_\odot$ \citep{AEF}, and for Cl\,J0023+0423B we
take the average of the projected and annular mass estimates within
714 kpc \citep{PLO-II}, and apply a small correction to scale the mass
to $R_{200}=614$ kpc.  This gives a mass of 
$2.3\pm1.2 \times10^{14}M_\odot$, where the uncertainty includes an
estimate of the systematic uncertainty, taken to be the difference between the
projected and annular mass measurements.
The results are shown in the top panel of
Fig.~\ref{fig:FZM}, and again show a strong
scatter  -- which likely exceeds any redshift evolution within the sample.
We limit the  
evolution to be a factor $\sim 4$ out to $z\sim 0.8$. However, we have not
accounted for systematic uncertainties in the mass estimates, which can
be important because they are not all calculated using the same technique.  
In fact, \citet{FZM} claimed to find a stronger
trend in normalised SFR with redshift; this is due to the use of mass
estimates based on a simple scaling of velocity dispersion or X-ray
luminosity.  In particular, their dynamical mass estimate (based on a
simple scaling of velocity dispersion) for Cl\,J0023+0423B is a
factor $\sim 2$--$4$ lower than the one we have derived from \citet{PLO-II};
adopting this mass gives the diamond point in Fig.~\ref{fig:FZM} and implies
stronger evolution out to $z\sim 0.8$.

On the basis of these five clusters it is not possible to draw strong
conclusions about the amount of evolution in the
mass-normalised SFR in clusters.  The data
suggest a large range  in the total
SFR per unit cluster mass at a fixed redshift, with limits on
the evolution with redshift which are comparable
to or even stronger than that seen in  the general field
\citep[e.g.][]{L96,Wilson+02}.  However, any evolutionary trend is
particularly sensitive to the mass estimate for Cl\,0023J+0423B,
and would have to ignore the  low SFR in AC\,114.  Although the results
are intriguing, a larger sample of clusters is clearly needed.
Furthermore, the sensitivity of H$\alpha$ to instantaneous star
formation may mean that tests such as the above are strongly affected
by small changes in the recent dynamic history of the cluster, for example.
Averaging over larger cluster samples (or using indicators of star formation
over longer time-scales) will allow us to overcome this sensitivity.

\subsection{Galaxy transformations?}\label{sec-passive}

A possible explanation for the observed trends with local density and
radius is that interactions with the surrounding environment transform
galaxies from one type to another.  Below we discuss specific
observations which support this hypothesis and help to constrain the
underlying mechanisms.

\subsubsection{The shape of the H$\alpha$ luminosity function}

We found that the distribution of H$\alpha$ strengths, for those
galaxies detected in H$\alpha$, is not strongly dependent on
environment.  A similar phenomenon is observed at low redshift, both in
terms of galaxy SFR \citep{2dfsdss} and colours \citep{BB04}.  Since
the fraction of emission line galaxies depends strongly on local
environment, this means that 
any transformation in galaxy properties must occur on a short
time-scale, so that a substantial population of galaxies with low but
present star formation is never dominant.  
For example, if transformations occur at a uniform rate with redshift,
and are due to a decline in star formation over a time-scale of
$\gtrsim 1$ Gyr, the fraction of galaxies in a
transition state at any given epoch would be $1/13.7=7.3$ per cent, assuming
a Hubble time of 13.7 Gyr.  Although this is a small fraction of the
total population, in dense environments (where few galaxies are forming
stars at any rate) it would represent a large
proportion of those galaxies detected in H$\alpha$.
At $\Sigma>90$
Mpc$^{-2}$, only $\sim 35$ per cent of galaxies are detected in
H$\alpha$ and, in the above model, we would expect $\sim 21$ per cent
of them to have abnormally low SFR; this would easily be detectable as
a change in the shape of the H$\alpha$ luminosity function.  Therefore,
we suggest that the time-scale for truncating star formation must be
$<1$ Gyr.
An alternative explanation
is that the transformations occurred at much earlier times; however,
the strong evolution in the fraction of star forming galaxies relative
to $z=0$ clusters makes this seem unlikely.

\subsubsection{Passive spiral galaxies}

The presence of spiral galaxies with little or no star formation,
which we call ``passive spirals'', are the most direct evidence we have that
galaxies may transform from one morphological type to another via
changes to the SFR \citep[e.g.][]{C+98,P+99,lowlx-spectra}.
Recently, \citet{Goto_passive}
have suggested, using SDSS data, that such galaxies are associated with the
infall regions of clusters.
Of the 63 spectroscopic member galaxies in Cl\,0024 with spiral
morphologies (classification $3$--$6$) as determined from {\it HST}
data, 22 ($35 \pm7$ per cent) 
are undetected in \halpha\ emission.
This is a much larger
fraction than the 0.28 per cent found in the local field
\citet{Goto_passive}, and consistent with the $\sim 20$ per cent
observed in distant clusters \citep{P+99}.
Most of the passive galaxies in our sample are actually detected in
\halpha\ {\it absorption} over the full spatial extent, and two
are post-starburst galaxies (see below).
None of the galaxies are detected
spectroscopically in \oii\ \citep{Czoske_cat}, and they are
found throughout the cluster region, with no
preference for either the centre or the outskirts.  
These galaxies have a range of colours, though most are quite red, and only 
8/22 (36 per cent) are bluer than $(B-R)_c=2.0$.  

This supports a scenario in which environmentally--induced changes to
the SFR are at least partially independent from morphological changes
and, in particular, that star formation is the more sensitive of the
two.  It is possible that some morphological transformation occurs on a
longer time-scale, by the fading of a disk in which star formation has
stopped \citep{Bekki02}.
An alternative possibility is that the truncation of star formation and
morphology is driven by different physical mechanisms.  This is
supported by our results that show the truncation of star formation may
take place at large radii, $\gtrsim 1$ Mpc, while the fraction of
spiral and irregular galaxies remains constant beyond $\sim 0.5$ Mpc.
This may suggest that the SFR responds quickly when galaxies reach the
outskirts of clusters, while another
physical mechanism(s), such as harassment \citep{harass},
may play a role in transforming the morphologies
later on when the galaxies approach further inside the cluster core.
If the decline of SFR was more gradual, the trend with
density would be less sharp, and the distinction from the morphological
trend would be less clear.

\subsubsection{Galaxies with strong H$\delta$ absorption}

\begin{figure}
\leavevmode \epsfysize=8.5cm \epsfbox{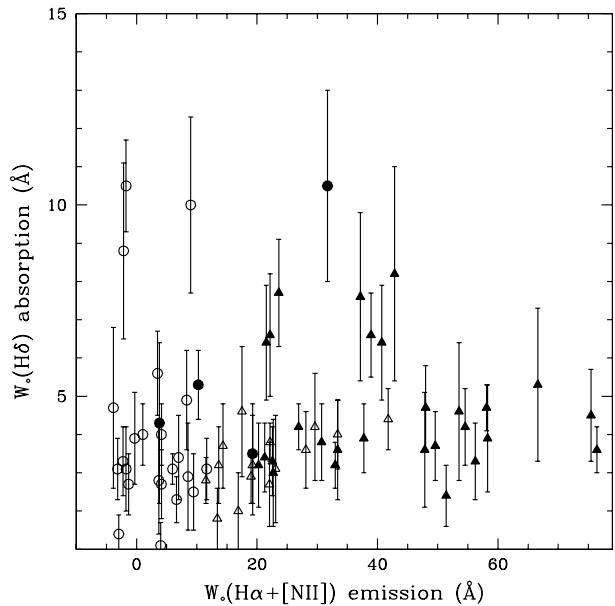}
\caption{The H$\delta$ absorption line equivalent width, from the
spectroscopic sample of \citet{Czoske_cat}, is shown as a function of
\ewhanii, for spectroscopically-confirmed cluster members.  {\it
Triangles} are H$\alpha$ detections, while {\it circles} are
non-detections.  {\it Open
symbols} have \oii\ equivalent widths $<10$~\AA, while {\it filled
symbols} have detectable emission with $>10$~\AA.  1-$\sigma$ error
bars are shown on H$\delta$ only, for clarity.
\label{fig:EA}}
\end{figure}

Finally, we consider another unusual class of galaxy,
comprised of those with atypically strong H$\delta$ absorption.
In particular, the
existence of such galaxies without the expected emission lines arising
from ionized gas have been invoked as evidence for a recent truncation
of star formation, perhaps preceded by a starburst
\citep{DG82,CS87,P+99,tomo-EA1}.  In Fig.~\ref{fig:EA} we show the
correlation between \ewhanii\ and $W_\circ($H$\delta)$,
for spectroscopically confirmed cluster members.  The H$\delta$
measurements are taken 
from the spectroscopic catalogue of \citet{Czoske_cat}; this line is
measured only for those galaxies in which it is particularly strong.  
Of the 243 spectroscopic members,
9 ($4 \pm 1$ per cent) have $W_\circ($H$\delta)>4$~\AA\ and no detectable
emission, approximately consistent or slightly larger than the fraction
in clusters at similar redshifts
\citep[e.g.][]{P+99,TranEA}.
All of these galaxies are located $>3$ arcmin ($>1$ Mpc) from the
cluster centre.   Unfortunately, the interpretation of these galaxies 
is not straightforward, as a simple cut in $W_\circ($H$\delta)$ is
insufficient to constrain the star formation history without ambiguity.
However, it is clear that the data are at least consistent with a
substantial population of post--starburst galaxies.  Since the duty cycle
of such galaxies is likely to be short, even a small population can be
indicative of a more widespread phenomenon.  The existence of these galaxies,
and the passive spiral galaxies noted above, are good evidence that
short time-scale transformations are taking place in the field around
Cl\,0024.  However, there are too few to pinpoint any particular
environment as the cause.

In addition, we see a very small number of galaxies with
$W_\circ($H$\delta)>4$~\AA\ and \ewoii\ $<10$~\AA, but which exhibit
strong H$\alpha$ emission, \ewhanii\ $\geq 30$~\AA\ (these are some
of the galaxies lying below the correlation in Fig.~\ref{fig:haoii}).
These apparently post-starburst galaxies are likely to be highly-obscured,
starbursts (where the [O{\sc ii}] emission is suppressed
by dust) similar to those discussed by \citet{Smail-radio} and \citet{PW00}.

\subsubsection{Summary}

All the above evidence implies that any transformations occurring in the
galaxy star formation rate must take place on relatively short time-scales,
in a manner that is independent of the morphological transformation.    
The fact that the fraction
of star--forming galaxies is reduced so far from the cluster centre
($\gtrsim 1$ Mpc) means that ram pressure stripping of the cold gas
through interaction with the intracluster medium \citep{GG} is
unlikely, although a significant fraction of galaxies at this radius
may have already passed through the cluster core \citep{MSSS,GKG}.  A second
possibility is that galaxy--galaxy interactions induce transformations
\citep[e.g.][]{harass,HO00,Lambas}, although the fact that the
morphological transformation appears to be decoupled from the decline
in star formation may prove a challenge for this interpretation.
A final possibility are the so-called strangulation models \citep{LTC,infall}
in which a purported halo of hot gas is stripped from galaxies in dense
environments, leading to a gradual winding down of star formation as
the remaining cold, disk gas is consumed.  
Although this may be effective in relatively low density environments,
the time-scale for the truncation of star formation is typically
$\gtrsim 1$ Gyr, which may be too long to accommodate our observations
that the shape of the H$\alpha$ luminosity function is insensitive to
environment.

The fact that the correlation between galaxy properties and local
density is already in place in this cluster at $z=0.4$, in a sense that
is qualitatively similar to that observed at $z=0$, may mean that the
predominant transformation mechanism occurs before galaxies are
accreted into clusters.

\section{Conclusions}\label{sec-conc}

We have used deep, panoramic multicolour imaging of Cl\,0024.0+1652 to
trace the galaxy population from the cluster core to the outskirts.
We find the following:

\begin{itemize}
\item[1.] The correlation between galaxy star formation rate and
local environment 
 is already well established in Cl\,0024
 ($z\sim 0.4$).  The fraction of galaxies detected in H$\alpha$ is a strong
  function of local projected density and cluster-centric radius.  The
  sense and magnitude of this trend is qualitatively similar to that
  seen at low redshift \citep{2dfsdss}.
\item[2.] The shape of the H$\alpha$ luminosity function
  is generally independent of environment, and consistent
  with the $z\sim 0.7$ field luminosity function of \citet{TMLC}.
  However, we detect a factor $\sim 3$ brightening in the
  characteristic luminosity  relative to lower
  redshift clusters  and the $z\sim 0.24$ field
  \citep{A2390_BM,C+01,A1689,Umeda03}.
\item[3.] The cluster core has a significant population ($35\pm7$ per cent)
  of spiral galaxies without H$\alpha$ emission, consistent with other
  clusters at similar redshifts \citep{P+99}.
  There is also a small population of post-starburst galaxies in the
  outskirts of the cluster.  
\item[4.] The fraction of galaxies with significant star formation 
  depends on environment far ($\gtrsim 1$ Mpc) from the cluster,
  while the fraction of spiral and irregular galaxies is only sensitive
  to environment within the inner $\sim 0.5$ Mpc.  This suggests that the
  two correlations may not entirely arise from the same underlying
  physical mechanisms.
\item[5.] Through a comparison with other studies of H$\alpha$ emission
  in clusters out to $z=0.8$, we find that the total SFR within
  $0.5R_{200}$ and the SFR normalised to total cluster mass both show a
  large cluster-to-cluster variation with no detectable dependence on
  redshift, although it would be possible for the latter to evolve
  as strongly as $(1+z)^6$ (given the large scatter).
  However, this is  dependent on the uncertain mass measurement of the
  most distant cluster.
\end{itemize}

We conclude that the decrease in star formation activity in dense
environments is likely due to the decline in SFR on relatively short
time-scales, before mechanisms such as galaxy harassment transform the
morphology of the galaxy.  The actual mechanism responsible for
truncating the star formation may occur even before galaxies are accreted
into clusters.

\section*{Acknowledgements}

We thank Oliver Czoske for providing his spectra for comparison
and Tommaso Treu for providing the reduced HST images.
We acknowledge the Suprime-Cam team for allowing us to use the
narrow-band filter, NB$_{912}$, prior to open to the public.
We thank Rose Finn for helpful discussions.
We also acknowledge the SDF team for allowing us to use their 
data as a control field to correct for the field
contamination in our analysis.
This work is based on data collected at Subaru
Telescope, which is operated by the National Astronomical Observatory of
Japan, and was financially supported in part by a Grant-in-Aid for the
Scientific Research (No.\, 15740126) by the Japanese Ministry of Education,
Culture, Sports, Science and Technology.
TK acknowledges the
Physics Department at University of Durham for their kind
hospitality during the course of this work.  Similarly, MLB thanks the
NAOJ for their hospitality.  MLB and RGB acknowledge financial support
from PPARC Research and Senior Fellowships,
respectively. IRS
acknowledges support from the Royal Society.

\setlength{\bibhang}{2.0em}

\bibliography{ms}

\end{document}